\documentclass[preprint,12pt]{elsarticle}

\usepackage{graphicx}

\usepackage{amssymb}
\usepackage{amsthm}

\usepackage{siunitx}  
\usepackage{amsmath}  
\usepackage{mathtools}  
\DeclareFontFamily{\encodingdefault}{\ttdefault}{\hyphenchar\font=`\-} 
\usepackage{listings}  
\usepackage{bm}

\newcounter{bla}

\journal{Computer Physics Communications}

\begin{document}

\begin{frontmatter}



\title{PyLlama: a stable and versatile Python toolkit for the electromagnetic modeling of multilayered anisotropic media}


\author[a]{M\'{e}lanie M. Bay}
\author[a]{Silvia Vignolini}
\author[b]{Kevin Vynck\corref{author}}

\cortext[author] {Corresponding author.\\\textit{E-mail address:} kevin.vynck@univ-lyon1.fr; \textit{Current address:} Institute of Light and Matter (ILM), CNRS, University Lyon 1, 69100 Villeurbanne, France}
\address[a]{Department of Chemistry, University of Cambridge, Lensfield Road, Cambridge, CB2 1EW, UK}
\address[b]{Laboratoire Photonique Num\'erique et Nanosciences (LP2N), CNRS, IOGS, Univ. Bordeaux, 33400 Talence, France}

\begin{abstract}




PyLlama is a handy Python toolkit to compute the electromagnetic reflection and transmission properties of arbitrary multilayered linear media, including the case of anisotropy. Relying on a $4 \times 4$-matrix formalism, PyLlama implements not only the transfer matrix method, that is the most popular choice in existing codes, but also the scattering matrix method, which is numerically stable in all situations (e.g., thick, highly birefringent cholesteric structures at grazing incident angles). PyLlama is also designed to suit the practical needs by allowing the user to create, edit and assemble layers or multilayered domains with great ease. In this article, we present the electromagnetic theory underlying the transfer matrix and scattering matrix methods and outline the architecture and main features of PyLlama. Finally, we validate the code by comparison with available analytical solutions and demonstrate its versatility and numerical stability by modelling cholesteric media of varying complexity. A detailed documentation and tutorial are provided in a separate user manual. Applications of PyLlama range from the design of optical components to the modelling of polaritonic effects in polar crystals, to the study of structurally coloured materials in the living world.

\end{abstract}


\end{frontmatter}



{\bf PROGRAM SUMMARY}

\begin{small}
\noindent
{\em Program Title:} PyLlama -- Python Toolkit for the Electromagnetic Modelling of Multilayered Anisotropic Media \\
{\em Developer’s repository link:} \texttt{https://github.com/VignoliniLab/PyLlama} \\
{\em Licensing provisions:} GPLv3          \\
{\em Programming language:} Python                                \\
{\em Supplementary material:} User guide and tutorials at \texttt{https://pyllama.readthedocs.io/}           \\
%
{\em Nature of problem:}
Computation of the optical reflection and transmission coefficients of arbitrary multilayered linear media, composed of an arbitrary number of layers, possibly mixing isotropic and anisotropic, absorbing and non-absorbing materials, for linearly or circularly polarized light.\\
{\em Solution method:}
Implementation of both the transfer matrix method (faster) and the scattering matrix method (more robust) relying on a $4 \times 4$ matrix formalism.\\
{\em Additional comments including Restrictions and Unusual features:}
Integration of a physical model to handle cholesteric structures, blueprint for the integration of user-created custom systems, hassle-free export of spectra for non-programmers even for complex and/or custom systems. External routines include: Numpy~\cite{1}, Scipy~\cite{2}, as well as Sympy~\cite{3} (optional).
   \\

\end{small}


\section{Introduction}
\label{sec:introduction}

Multilayered media made of anisotropic materials are widespread in nature~\cite{Kinoshita2008} and in nanotechnologies, especially for optoelectronic~\cite{steiner2004semiconductor}, optical and photonic applications~\cite{McCall2014, mackay2019electromagnetic}. For instance, chiral nematic (or cholesteric) structures consist in birefringent units arranged into periodic helicoidal architectures, which can selectively reflect circularly-polarised light in a specific wavelength range~\cite{Oseen1933, DeVries1951, Belyakov1979, Werbowyj1984, Sixou1991, Frka-Petesic2019}. In living organisms, these structures are responsible for the vibrant blue of \textit{Pollia condensata} fruits~\cite{Vignolini2012} and the shiny green of \textit{Scarabeid} beetles~\cite{Carter2016}. They can also be fabricated through self-assembly mechanisms from bio-compatible materials, such as cellulose nanocrystals~\cite{Parker2016}, cellulose derivatives~\cite{Gray1994} or their composites~\cite{Saraiva2020, Boott2020}, with applications in photonic pigments, color printing and optical sensors~\cite{Parker2016, Bardet2015, Yi2019, Zhao2019, Boott2020}. In optical engineering, multilayered anisotropic media have been used to realize linear polarizers, waveplates and birefrigent filters for laser and telecommunication technologies, imaging systems, displays or gas sensing~\cite{McCall2014, Yeh2009}. More recently, they opened new perspectives for subdiffraction wave focusing~\cite{Dai2015, Li2015} thanks to surface phonon polaritons~\cite{dai2014tunable} and low-loss surface wave guiding~\cite{Pulsifer2013, Mackay2019}.

Analytical solutions for the optical properties of anisotropic multilayers are only available in specific situations (e.g., for periodic cholesterics, at normal incidence~\cite{DeVries1951, Chandrasekhar1968} or at oblique incidence with certain restrictions~\cite{Belyakov1979, Fergason1966}). A rigorous and general theoretical framework was established with the seminal contributions of Billard~\cite{Billard1966}, Teitler and Henvis~\cite{Teitler1970}, and Berreman~\cite{Berreman1972}, who expressed the electromagnetic problem as a $4 \times 4$ matrix ordinary differential equation. $4 \times 4$ matrix formalisms experienced many developments over the years~\cite{Yeh1979, LinChung1984, Ko1988, Eidner1989, Oldano1989, Schubert1996, schubert1999explicit, Stallinga1999}, including faster algorithms~\cite{Palto2001} and a correct treatment of degeneracies causing singularities~\cite{Xu2000, Passler2017}. Besides commercial solutions, such as the powerful program WVASE~\cite{WVASE} dedicated to ellipsometry analysis~\cite{schubert2004infrared}, freely-available and open-access codes~\cite{Castany2016, Passler2017code} generally exploit the elegant transfer matrix method~\cite{Mackay2020} to propagate the solution from layer to layer, as originally proposed by Berreman~\cite{Berreman1972}. Unfortunately, such an approach can become numerically unstable for large systems in presence of evanescent modes due to the coexistence of exponentially decaying and growing waves (in the forward propagating direction) in the multilayer. This shortcoming was addressed by Ko and Sambles~\cite{Ko1988}, who suggested to use a scattering matrix to treat all evanescent modes as forward or backward decaying waves. Scattering matrix formalisms have gained popularity in the framework of the rigorous coupled wave analysis (RCWA) for periodically-corrugated multilayered media~\cite{cotter1995scattering, li1996formulation, whittaker1999scattering, tikhodeev2002quasiguided, liscidini2008scattering} to avoid numerical instabilities. Some freely-available RCWA codes~\cite{hugonin2021reticolo, liu2012s4, germer2021pyscatmesh} can be used for uncorrugated multilayered anisotropic media, yet with an unnecessary complexity. The scattering matrix method has been implemented recently for the study of cholesterics but at normal incidence only~\cite{Kragt2019}.

Here, we present a freely-available and open-source Python toolkit named ``PyLlama''~\footnote{\underline{Py}thon toolkit for mu\underline{l}ti\underline{l}ayered \underline{a}nisotropic \underline{m}edi\underline{a}} that implements both the transfer matrix and the scattering matrix methods to compute the reflection and transmission properties of arbitrary (linear and uncorrugated) multilayered media in all possible situations. While the transfer matrix method proves being faster, the scattering matrix method ensures numerical stability in all cases. PyLlama can thus deal with systems composed of an arbitrary number of layers of any thicknesses, mixing isotropic and anisotropic, absorbing and non-absorbing materials, studied at all possible incident angles and wavelengths for both linear and circular polarizations. Surface waves may be excited by modelling the Otto-Kretschmann configurations~\cite{sambles1991optical, Passler2017}.

In addition to robustness, we designed the code to be convenient to use and included tools that are particularly suitable for the modelling of cholesteric liquid crystals. First, layers are treated as building blocks. The optical calculations occur internally at the layer level and the user can write their own routines, in form of new classes, to pile up layers into stacks according the parameters of their choice without having to handle optical parameters. Second, the code is organised in modules. Periodic structures can be easily combined into master structures with sub-periodicities, and different pre-defined structures can be stacked onto each other, in order to enable the modelling of complex situations, such as beetle cuticles made of a combination of cholesterics and absorbing layers~\cite{Carter2016, Mendoza-Galvan2018} or stacks of cellulose nanocrystal cholesteric structures separated by a nematic layer~\cite{Fernandes2017}. Third, once the user has defined which structures they need to model, a level of automation has been added to allow for the calculation of spectra and the export in MATLAB and/or Pickles with only a few lines of code. Lastly, more specific to cholesterics, we incorporated a physical model~\cite{Frka-Petesic2019} that describes the cholesteric helicoid in various situations (choice of handedness, tilt, vertical compression of tilted helicoids). In particular, this enables the optical modelling (at normal and oblique incidences) of distorted cholesterics for which the distortion impacts the angular response, polarisation selectivity, and overall reflection spectrum of the structure~\cite{Dreher1973, Frka-Petesic2019, Kragt2019, Chan2019, Boott2020}.

PyLlama addresses a need from the scientific community for a simple, robust and flexible program, with many possibilities for future developments and improvements thanks to its object-oriented organisation and detailed documentation.

The following sections are organized as follows. Section~\ref{sec:theory} describes the electromagnetic theory underlying the $4 \times 4$ matrix formalism, as well as the transfer matrix and scattering matrix methods to propagate the solution throughout the multilayer stack. Section~\ref{sec:code_structure} outlines the architecture and main features of PyLlama. Several practical examples of optical computations and comparisons with analytical solutions are given to illustrate the method and show its versatility. The performance and stability of the transfer matrix and scattering matrix methods are compared in Section~\ref{sec:comp_TM_SM} on the example of a cholesteric structure. Finally, we provide concluding remarks in Section~\ref{sec:concluding_remarks}. The Appendices contain details about the numerical analysis of the so-called partial waves in optical layers, a description of the discrete model for cholesterics, and all the material and wave parameters used to construct the figures.

\section{Theory}
\label{sec:theory}

\subsection{Description of the problem}
\label{sec:system}

\begin{figure}
  \centering
    \includegraphics[width=\textwidth]{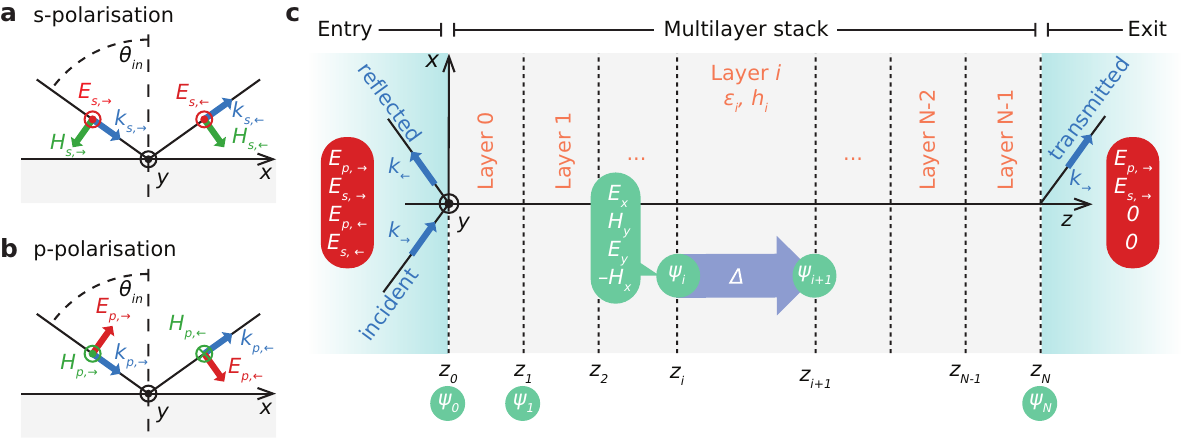}
	\caption{a, b) Schematic of an incident and a reflected $s$-polarised waves (a) or $p$-polarised waves (b) upon the stack. The plane of incidence is the $(x,z)$ plane. The wavevectors $k_\text{i}$ and $k_\text{r}$ form an angle $\theta$ with the $z$-axis. c) Schematic of the multilayer stack bounded by the entry and exit half-spaces. This schematic clarifies the notations used in the main text.}
	\label{fig:stack}
\end{figure}

We consider a multilayer stack composed of $N$~layers perpendicular to the $z$~axis and translationally-invariant in the $(xy)$~plane, see Fig.~\ref{fig:stack}. The layers are indexed from $0$ to $N-1$~to match Python's indexing convention. Each layer $i$ is bounded by two interfaces at $z_i$ and $z_{i+1}$, with an arbitrary thickness $h_i$, and described by an arbitrary permittivity tensor $\bm{\epsilon}_i$, which may contain complex values to describe absorbing materials. The multilayer stack is enclosed between two semi-infinite isotropic media called ``entry'' and ``exit'' with refractive indices $n_\text{entry}$ and $n_\text{exit}$, respectively.

The multilayer stack is illuminated from the entry half-space by a planewave at frequency $\omega$ propagating in the $(xz)$-plane at an angle $\theta_\text{in}$ from the surface normal $z$. The wavevector $\mathbf{k}$ of the incident planewave is thus given by
\begin{equation}
\label{eq:wavevector}
\mathbf{k} = 
\begin{bmatrix*}
k_x \\
k_y \\
k_z
\end{bmatrix*}
= k_0
\begin{bmatrix*}
K_x \\
K_y \\
K_z
\end{bmatrix*},
\end{equation}
with $k_0 = \frac{2 \pi}{\lambda}$. Here, $K_x = n_\text{entry} \sin({\theta_\text{in}})$ and $K_y = 0$ are constant throughout the multilayer stack.

In the isotropic half-spaces and layers, planewaves can be decomposed into s and p polarisations, see Fig.~\ref{fig:stack}(a,b). The s-polarised wave has its electric field perpendicular (\textit{s} from the German \textit{senkrecht}, perpendicular) to the plane of incidence $(xz)$, along $y$, and the p-polarised wave has its electric field in the plane of incidence $(xz)$. Of particular relevance for helicoidal structures, the waves can also be described in a circular polarisation basis denoted as R and L for the right and left-handed polarisations, respectively. 
Our objective here is to compute the complex reflection and transmission coefficients of the multilayer stack, defined as $r_{kj}$ and $t_{kj}$, where $j$ indicates the polarisation of the incident wave, either in the linear or the circular polarisation basis, and $k$ indicates the polarisation of the reflected or transmitted wave, respectively. Note that for multilayers containing anisotropic layers, the cross-terms in the linear polarisation basis ($r_\text{ps}$, $r_\text{sp}$, $t_\text{ps}$ and $t_\text{sp}$) can be non-zero. The reflectance and the transmittance can then be calculated respectively as $R_{kj} = |r_{kj}| ^ 2$ and $T_{kj} = |t_{kj}| ^ 2 \cos(\theta_\text{in}) / \cos(\theta_\text{out})$~\cite{Yeh1988}, where $\theta_\text{out} = \text{asin} \left[ n_\text{entry} \sin (\theta_\text{in})/n_\text{exit}  \right]$ is the angle of the planewave transmitted in the exit half-space.

\subsection{$4 \times 4$ matrix formalism}
\label{sec:berreman_matrix}

Maxwell's equations set relationships between the 3~components of the electric field and the 3~components of the magnetic field of a lightwave in a given medium, which, in a matrix form, read
\begin{equation}
\label{eq:berreman_maxwell}
\begin{bmatrix*}[c]
0 & 0 & 0 & 0 & -\frac{\partial}{\partial z} & \frac{\partial}{\partial y} \\
0 & 0 & 0 & \frac{\partial}{\partial z} & 0 & -\frac{\partial}{\partial x} \\
0 & 0 & 0 & -\frac{\partial}{\partial y} & \frac{\partial}{\partial x} & 0 \\
0 & \frac{\partial}{\partial z} & -\frac{\partial}{\partial y} & 0 & 0 & 0 \\
-\frac{\partial}{\partial z} & 0 & \frac{\partial}{\partial x} & 0 & 0 & 0 \\
\frac{\partial}{\partial y} & -\frac{\partial}{\partial x} & 0 & 0 & 0 & 0 \\
\end{bmatrix*}
\begin{bmatrix*}[c]
E_x \\
E_y \\
E_z \\
H_x \\
H_y \\
H_z \\
\end{bmatrix*}
=
\frac{1}{c} \frac{\partial}{\partial t}
\begin{bmatrix*}[c]
D_x \\
D_y \\
D_z \\
B_x \\
B_y \\
B_z \\
\end{bmatrix*}.
\end{equation}
Introducing the permittivity $\bm{\epsilon}$, permeability $\bm{\mu}$ and optical activity $\bm{\rho}$ and $\bm{\rho}'$ tensors of the material to relate the $\mathbf{D}$ and $\mathbf{B}$ fields to the $\mathbf{E}$ and $\mathbf{H}$ fields, applying translational-invariance in the $x$ and $y$ directions and considering harmonic fields (the $\exp \left( -i \omega t\right)$ convention is used hereafter), Eq.~\eqref{eq:berreman_maxwell} can be rewritten as
\begin{equation}
\label{eq:bereq-left-side}
\begin{bmatrix*}[c]
0 & 0 & 0 & 0 & -\frac{\partial}{\partial z} & 0 \\
0 & 0 & 0 & \frac{\partial}{\partial z} & 0 & -i K_x \\
0 & 0 & 0 & 0 & i K_x & 0 \\
0 & \frac{\partial}{\partial z} & 0 & 0 & 0 & 0 \\
-\frac{\partial}{\partial z} & 0 & i K_x & 0 & 0 & 0 \\
0 & -i K_x & 0 & 0 & 0 & 0 \\
\end{bmatrix*}
\begin{bmatrix*}[c]
E_x \\
E_y \\
E_z \\
H_x \\
H_y \\
H_z \\
\end{bmatrix*}
=
- \frac{i \omega}{c}
\begin{bmatrix*}[c]
\epsilon_{xx} & \epsilon_{xy} & \epsilon_{xz} & \rho_{xx} & \rho_{xy} & \rho_{xz} \\
\epsilon_{yx} & \epsilon_{yy} & \epsilon_{yz} & \rho_{yx} & \rho_{yy} & \rho_{yz} \\
\epsilon_{zx} & \epsilon_{zy} & \epsilon_{zz} & \rho_{zx} & \rho_{zy} & \rho_{zz} \\
\rho'_{xx} & \rho'_{xy} & \rho'_{xz} & \mu_{xx} & \mu_{xy} & \mu_{xz} \\
\rho'_{yx} & \rho'_{yy} & \rho'_{yz} & \mu_{yx} & \mu_{yy} & \mu_{yz} \\
\rho'_{zx} & \rho'_{zy} & \rho'_{zz} & \mu_{zx} & \mu_{zy} & \mu_{zz} \\
\end{bmatrix*}
\begin{bmatrix*}[c]
E_x \\
E_y \\
E_z \\
H_x \\
H_y \\
H_z \\
\end{bmatrix*}.
\end{equation}
Expressing the normal ($z$) components as a function of the tangential ($x$, $y$) components, taking out the third and six rows and rearranging the matrices finally leads to the famous $4 \times 4$ matrix ordinary differential equation~\cite{Billard1966,Teitler1970,Berreman1972}
\begin{equation}\label{eq:ODE}
\frac{\partial \bm{\Psi}(z)}{\partial z}
=
\frac{i \omega}{c}
\bm{\Delta}(z)
\bm{\Psi}(z),
\end{equation}
where $\bm{\Psi}=[E_x, H_y, E_y, -H_x]^\text{T}$ is a vector describing the field at position $z$.\footnote{The field can indifferently be represented by the vector $\bm{\Psi} = [E_x, H_y, E_y, -H_x]^\text{T}$ as in Refs.~\cite{Berreman1972, Oldano1989, Palto2001, Passler2017, Kragt2019}, or by $\bm{\Psi} = [E_x, E_y, H_x, H_y]^\text{T}$ as in Ref.~\cite{Castany2016}.}

The elements of the matrix $\bm{\Delta}$ depend on the wavevector constant component $K_x$ and on the material permittivity, permeability and optical activity. When the material is non-magnetic and non-optically active, the matrix $\bm{\Delta}(z)$ explicitly reads
\begin{equation}
\label{eq:delta}
\bm{\Delta}(z)
=
\begin{bmatrix*}[c]
- K_x \frac{\epsilon_{zx}}{\epsilon_{zz}}
	& 1 - \frac{K_x^2}{\epsilon_{zz}}
	& - K_x \frac{\epsilon_{zy}}{\epsilon_{zz}}
	& \frac{K_x}{\epsilon_{zz}} \\
\epsilon_{xx} - \frac{\epsilon_{xz} \epsilon_{zx}}{\epsilon_{zz}}
	& - K_x \frac{\epsilon_{xz}}{\epsilon_{zz}}
	& \epsilon_{xy} - \frac{\epsilon_{xz} \epsilon_{zy}}{\epsilon_{zz}}
	& \frac{\epsilon_{xz}}{\epsilon_{zz}} \\
0 & 0 & 0 & 1 \\
\epsilon_{yx} - \frac{\epsilon_{yz} \epsilon_{zx}}{\epsilon_{zz}}
	& - K_x \frac{\epsilon_{yz}}{\epsilon_{zz}}
	& \epsilon_{yy} - K_x^2 - \frac{\epsilon_{yz} \epsilon_{zy}}{\epsilon_{zz}}
	& \frac{\epsilon_{yz}}{\epsilon_{zz}} \\
\end{bmatrix*},
\end{equation}
where the $z$-dependence is in the permittivity tensor $\bm{\epsilon}$.

\subsection{Propagation in a homogeneous layer}
\label{sec:one_layer}

Let us now consider a homogeneous layer $i$ located between two interfaces at positions $z_{i}$ and $z_{i+1}$, see Fig.~\ref{fig:stack}(c). The corresponding $4 \times 4$ matrix $\bm{\Delta}_i$ is then constant throughout the layer thickness and Eq.~\eqref{eq:ODE} directly leads to
\begin{equation}
\label{eq:integ_ODE}
\bm{\Psi}(z_{i+1}) = \exp \left( i k_0 h_i \bm{\Delta}_i \right) \bm{\Psi}(z_i)
= \mathbf{R}_i \bm{\Psi}(z_i)
\end{equation}
where $h_i=z_{i+1} - z_{i}$ is the layer thickness and $\mathbf{R}_i$ describes the propagation from $z_i$ to $z_{i+1}$.

Discarding the layer label $i$ for the sake of legibility, there are different ways to calculate the propagator $\mathbf{R}$:
\begin{enumerate}
\item Expanding $\mathbf{R}$ as an infinite sum, as $\exp \left( i k_0 h \bm{\Delta} \right) = 1 + i k_0 h \bm{\Delta} - \frac{1}{2} k_0^2 h^2 \bm{\Delta}^2 + ...$ as suggested by Berreman~\cite{Berreman1972}, or via Padé's decomposition as in Ref.~\cite{Castany2016}. This is directly implemented in the matrix exponential function \texttt{expm} in Python’s package Scipy (version 1.4.1) and in MATLAB (version R2019b).
\item Factorizing $\bm{\Delta} = \mathbf{P} \mathbf{Q}_{\bm{\Delta}} \mathbf{P}^{-1}$ where $\mathbf{Q}_{\bm{\Delta}}$ is a diagonal matrix containing the eigenvalues and $\mathbf{P}$ the eigenvectors of $\bm{\Delta}$. This leads to an eigendecomposition of $\mathbf{R}$ as $\mathbf{R} = \mathbf{P} \mathbf{Q} \mathbf{P}^{-1}$, with identical eigenvectors and eigenvalues straightforwardly related those of $\bm{\Delta}$, see below. This was also suggested by Berreman~\cite{Berreman1972} and is implemented in Passler and Paarmann's code~\cite{Passler2017, Passler2017code}.
\item Using an algorithm based on Sylvester's formulas applied to biaxial crystals, as proposed by Palto \textit{et al.}~\cite{Palto2001}, which is expected to converge faster and requires less calculation steps than Padé's decomposition.
\end{enumerate}

\begin{figure}
  \centering
    \includegraphics[width=\textwidth]{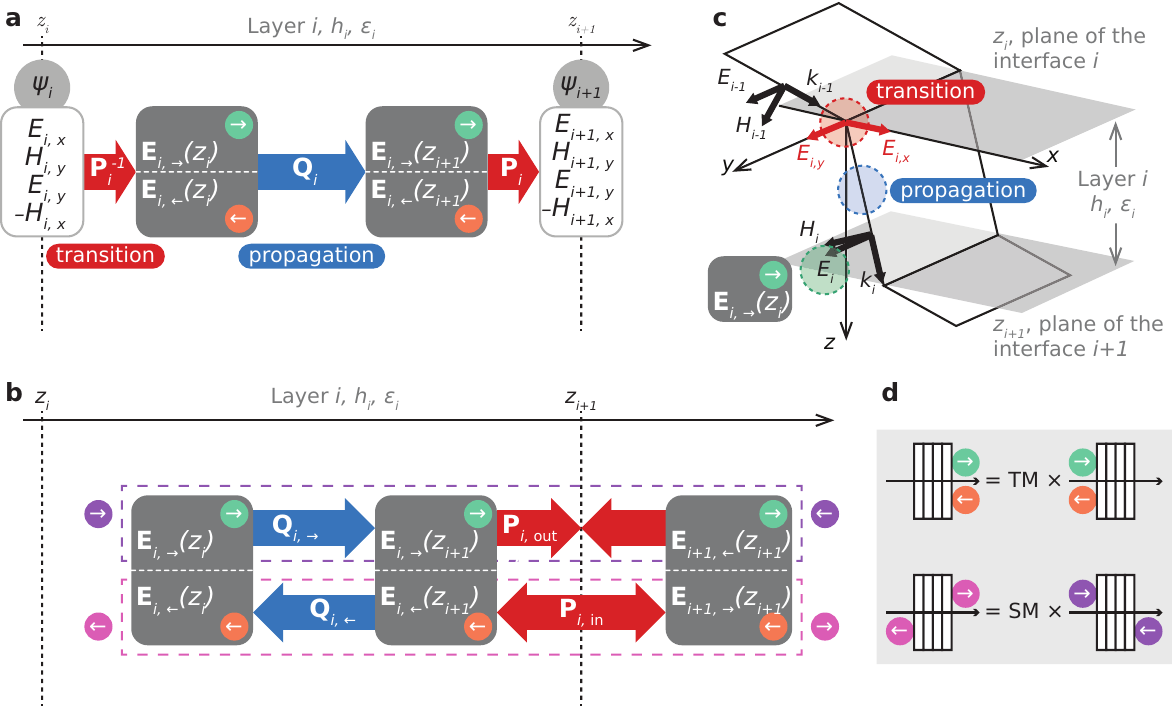}
	\caption{a, b) Schematic representations of the variables used to describe the light interaction with the layer $i$ a) with the transfer matrix method and b) with the scattering matrix method. Both methods are based on the phenomena of transition and propagation, which are schematised with wide red or blue arrows labelled with the corresponding matrices used in the equations. c) 3D representation of a layer and its interfaces. Transition occurs at the interface between two layers and propagation occurs within the thickness of the layer. d) Schematic representation of the transfer matrix method and the scattering matrix method.}
	\label{fig:SM_TM}
\end{figure}

In the second approach, which we will follow here, the fields in the layer are decomposed into four \textit{partial waves}~\cite{Yeh1988}. The eigendecomposition has a clear physical meaning, as sketched in Fig.~\ref{fig:SM_TM}a. The matrix $\mathbf{P}$ contains the eigenvectors $\mathbf{p}_j$ of $\bm{\Delta}$ as
\begin{equation}
\mathbf{P} =
\begin{bmatrix*}[c]
\mathbf{p}_0 & \mathbf{p}_1 & \mathbf{p}_2 & \mathbf{p}_3
\end{bmatrix*},
\end{equation}
where $\mathbf{p}_j = [E_{j,x}, H_{j,y}, E_{j,y}, -H_{j,x}]^\text{T}$ in $\mathbf{P}$ thus represent the partial wave field components in the homogeneous layer. The diagonal matrix $\mathbf{Q} = \exp(i k_0 h \mathbf{Q}_{\bm{\Delta}})$ depends on the corresponding eigenvalues $q_j$ of $\bm{\Delta}$ as
\begin{equation}
\mathbf{Q}
=
\begin{bmatrix*}[c]
e^{i k_0 h q_0} & 0 & 0 & 0 \\
0 & e^{i k_0 h q_1} & 0 & 0 \\
0 & 0 & e^{i k_0 h q_2} & 0 \\
0 & 0 & 0 & e^{i k_0 h q_3}
\end{bmatrix*}.
\end{equation}
It thus describes the coherent propagation of the partial waves in the layer, where the eigenvalues $q_j$ represent the $z$-components of the corresponding wavevectors $K_{j,z}$.

The four partial waves can generally be represented as a pair of waves travelling forward (towards $+z$, subscript $\rightarrow$) and a pair of partial waves that travel backward (towards $-z$, subscript $\leftarrow$). The two partial waves within each pair can then be identified according to their polarisation. In PyLlama, the partial waves are sorted as follows
\begin{itemize}
\item $j=0$: forward direction ($\rightarrow$), mostly polarised along the $x$ axis,
\item $j=1$: forward direction ($\rightarrow$), mostly polarised along the $y$ axis,
\item $j=2$: backward direction ($\leftarrow$), mostly polarised along the $x$ axis,
\item $j=3$: backward direction ($\leftarrow$), mostly polarised along the $y$ axis
\end{itemize}
but this is only one possibility~\footnote{In Ref.~\cite{Yeh1979}, the partial waves are sorted according to their polarisation first and their direction second (which would correspond to the order 0, 2, 1, 3, here).}. Details about the numerical analysis of the partial waves is provided in~\ref{sec:partial_waves}.

Let us emphasize that the sorting is quite irrelevant for obtaining the transfer and scattering matrices of the multilayer stack (when one does not wish to extract transfer and scattering matrices for subsets of the system), the only constraint being that the intermediate matrices used in the calculation be invertible. The sorting allows avoiding such situations.

\subsection{Transfer matrix method}
\label{sec:transfer_matrix}

The computation of $\mathbf{R}_i$ for layer $i$ allows us to relate, via Eq.~\eqref{eq:integ_ODE}, the tangential components of the electric and magnetic fields at the interface $z_i$ to those at the interface $z_{i+1}$.

Imposing the continuity of the field tangential components at each interface naturally leads to a generalization of the propagator for a multilayer stack of $N$ layers ($N+1$ interfaces) as~\cite{Berreman1972, Mackay2020}
\begin{equation}
\label{eq:berprod}
\bm{\Psi}_{N} = \mathbf{R} \bm{\Psi}_0,
\end{equation}
with
\begin{equation}
\label{eq:propagator_multistack}
\mathbf{R} = \overset{\curvearrowleft}{\prod_{i=0}^{N-1}} \mathbf{R}_i = \overset{\curvearrowleft}{\prod_{i=0}^{N-1}} \mathbf{P}_i \mathbf{Q}_i \mathbf{P}_i^{-1},
\end{equation}
and $\overset{\curvearrowleft}{\prod}_{i=0}^{N-1}$ is a left-side matrix product.\footnote{We define here the matrix $\mathbf{Q}_i$ similarly to Berreman~\cite{Berreman1972}, but its inverse ($\mathbf{Q}_i' = \mathbf{Q}_i^{-1}$) is sometimes used in the literature~\cite{Passler2017, Castany2016}; in this case, the left-side product in Eq.~\eqref{eq:berprod} should be replaced by a right-side product $\overset{\curvearrowright}{\prod}_{i=0}^{N-1} \mathbf{P}_i \mathbf{Q}_i' \mathbf{P}_i^{-1}$, leading to a propagator that is the inverse of the one considered here.}

It is usually more convenient to express the electromagnetic field in terms of s and p polarisations in the entry and exit isotropic half-spaces. For a planewave in an isotropic medium, the s- and p-polarised components of the electric and magnetic fields are related to the tangential components of the fields as~\cite{Yeh1988}
\begin{equation}
\begin{bmatrix*}[c]
E_x \\ H_y \\ E_y \\ -H_x
\end{bmatrix*}
=
\mathbf{L}
\begin{bmatrix*}[c]
E_{p,\rightarrow} \\ E_{s,\rightarrow} \\ E_{p,\leftarrow} \\ E_{s,\leftarrow}
\end{bmatrix*},
\label{eq:half-spaces}
\end{equation}
with
\begin{equation}
\mathbf{L} = 
\begin{bmatrix*}[c]
\cos(\theta) & 0 & \cos(\theta) & 0 \\
n & 0 & -n & 0 \\
0 & 1 & 0 & 1 \\
0 & n \cos(\theta) & 0 & - n \cos(\theta)
\end{bmatrix*}.
\end{equation}
Here, $n=\{n_\text{entry},n_\text{exit}\}$ and $\theta=\{\theta_\text{in},\theta_\text{out}\}$ for the entry and exit media, respectively. Combining the corresponding matrices, $\mathbf{L}_\text{entry}$ and $\mathbf{L}_\text{exit}$, at both ends of Eq.~\eqref{eq:berprod}, we can now define the \textit{transfer matrix} $\mathbf{T}$ of the multilayer stack relating the ingoing and outgoing fields at the first interface $z_0$ to the ingoing and outgoing fields at the last interface $z_N$,
\begin{equation}
\begin{bmatrix*}[c]
E_{p,\rightarrow}(z_N) \\ E_{s,\rightarrow}(z_N) \\ E_{p,\leftarrow}(z_N) \\ E_{s,\leftarrow}(z_N)
\end{bmatrix*}
=
\mathbf{T}
\begin{bmatrix*}[c]
E_{p,\rightarrow}(z_0) \\ E_{s,\rightarrow}(z_0) \\ E_{p,\leftarrow}(z_0) \\ E_{s,\leftarrow}(z_0)
\end{bmatrix*}.
\label{eq:full_transfer_matrix_1}
\end{equation}
with
\begin{equation}
\mathbf{T} = \mathbf{L}_\text{exit}^{-1} \mathbf{R} \mathbf{L}_\text{entry}.
\label{eq:full_transfer_matrix_2}
\end{equation}

The complex reflection and transmission coefficients are finally obtained as
\begin{alignat}{3}
\label{eq:reflco_tm}
r_{\text{pp}} & = & \dfrac{T_{30} T_{23} - T_{20} T_{33}}{T_{22} T_{33} - T_{32} T_{23}} \notag \\
r_{\text{ps}} & = & \dfrac{T_{31} T_{23} - T_{21} T_{33}}{T_{22} T_{33} - T_{32} T_{23}} \notag \\
r_{\text{sp}} & = & \dfrac{T_{20} T_{32} - T_{30} T_{22}}{T_{22} T_{33} - T_{32} T_{23}} \notag \\
r_{\text{ss}} & = & \dfrac{T_{21} T_{32} - T_{31} T_{22}}{T_{22} T_{33} - T_{32} T_{23}} \notag \\
t_{\text{pp}} & = & T_{00} + T_{02} r_{\text{pp}} + T_{\text{13}} r_{\text{sp}} \notag \\
t_{\text{ps}} & = & T_{01} + T_{02} r_{\text{ps}} + T_{\text{03}} r_{\text{ss}} \notag \\
t_{\text{sp}} & = & T_{10} + T_{12} r_{\text{pp}} + T_{\text{03}} r_{\text{sp}} \notag \\
t_{\text{ss}} & = & T_{11} + T_{12} r_{\text{ps}} + T_{\text{13}} r_{\text{ss}}
\end{alignat}
These coefficients can conveniently be assembled into a Jones matrix and expressed on the circular polarisation basis from~\cite{Yeh1988}
\begin{alignat}{3}
\begin{bmatrix*}[c]
r_{\text{RR}} & r_{\text{RL}} \\
r_{\text{LR}} & r_{\text{LL}}
\end{bmatrix*}
& = &
\left( \begin{bmatrix*}[c]
1 & 1 \\
i & -i
\end{bmatrix*} \right) ^ {-1}
\begin{bmatrix*}[c]
r_{\text{pp}} & r_{\text{ps}} \\
r_{\text{sp}} & r_{\text{ss}}
\end{bmatrix*}
\begin{bmatrix*}[c]
1 & 1 \\
-i & i
\end{bmatrix*} \label{eq:jones_circ_1} \\
\begin{bmatrix*}[c]
t_{\text{RR}} & t_{\text{RL}} \\
t_{\text{LR}} & t_{\text{LL}}
\end{bmatrix*}
& = &
\left( \begin{bmatrix*}[c]
1 & 1 \\
-i & i
\end{bmatrix*} \right) ^ {-1}
\begin{bmatrix*}[c]
t_{\text{pp}} & t_{\text{ps}} \\
t_{\text{sp}} & t_{\text{ss}}
\end{bmatrix*}
\begin{bmatrix*}[c]
1 & 1 \\
-i & i
\end{bmatrix*}
\label{eq:jones_circ_2}
\end{alignat}

\subsection{Scattering matrix method}
\label{sec:scattering_matrix}

The scattering matrix method was proposed as a solution to the numerical instabilities encountered by the transfer matrix method in certain situations, such as cholesteric liquid crystals~\cite{Ko1988}. The scattering matrix relates the waves coming in the multilayer stack from both sides to the waves coming out from it, in contrast with the transfer matrix which relates the waves at the first interface to the waves at the last interface, see Eq.~\eqref{eq:full_transfer_matrix_1}.

Following the transfer matrix formalism above, the transition from the set of partial waves in layer $i$ to those in layer $i+1$ can simply be expressed as
\begin{equation}
\label{eq:tm_to_sm}
\mathbf{P}_{i+1}
\begin{bmatrix*}[c]
\mathbf{E}_{i+1, \rightarrow} (z_{i+1}) \\
\mathbf{E}_{i+1, \leftarrow} (z_{i+1})
\end{bmatrix*}
=
\mathbf{P}_{i}
\mathbf{Q}_{i}
\begin{bmatrix*}[c]
\mathbf{E}_{i, \rightarrow} (z_{i}) \\
\mathbf{E}_{i, \leftarrow} (z_{i})
\end{bmatrix*}
\end{equation}
Knowing the travel direction and the $z$ coordinate at which the electric fields are evaluated enables us to determine whether they are ingoing or outgoing waves, as schematised on Fig.~\ref{fig:SM_TM}b. Equation ~\eqref{eq:tm_to_sm} can be rearranged with linear operations to dispatch the forward- and backward-propagating waves into ingoing and outgoing waves, leading to
\begin{equation}
\label{eq:split_eigen}
\begin{array}{rcl}
\mathbf{P}_{i,\text{out}} 
\begin{bmatrix*}[c]
\mathbf{E}_{i, \rightarrow} (z_{i+1}) \\
\mathbf{E}_{i+1, \leftarrow} (z_{i+1})
\end{bmatrix*}
& = &
\mathbf{P}_{i,\text{in}}
\begin{bmatrix*}[c]
\mathbf{E}_{i+1, \rightarrow} (z_{i+1}) \\
\mathbf{E}_{i, \leftarrow} (z_{i+1})
\end{bmatrix*}, \\

\mathbf{E}_{i, \rightarrow} (z_{i+1}) & = & \mathbf{Q}_{i,\rightarrow} \mathbf{E}_{i, \rightarrow} (z_{i}), \\

\mathbf{E}_{i, \leftarrow} (z_{i+1}) & = & \mathbf{Q}_{i,\leftarrow} \mathbf{E}_{i, \leftarrow} (z_{i}),
\end{array}
\end{equation}
where the matrices $\mathbf{P}_{i,\text{out}}$, $\mathbf{P}_{i,\text{in}}$, $\mathbf{Q}_{i,\rightarrow}$ and $\mathbf{Q}_{i,\leftarrow}$ are defined as
\begin{equation}
\begin{array}{rcl}
\mathbf{P}_{i,\text{out}}
& = & 
\begin{bmatrix*}[c]
\mathbf{p}_{i, 0} & \mathbf{p}_{i, 1} & -\mathbf{p}_{i+1, 2} & -\mathbf{p}_{i+1, 3} 
\end{bmatrix*}, \\
\mathbf{P}_{i,\text{in}}
& = &
\begin{bmatrix*}[c]
\mathbf{p}_{i+1, 0} & \mathbf{p}_{i+1, 1} & -\mathbf{p}_{i, 2} & -\mathbf{p}_{i, 3}
\end{bmatrix*}, \\
\mathbf{Q}_{i,\rightarrow}
& = &
\begin{bmatrix*}[c]
e^{i k_0 h q_{i, 0}} & 0 & 0 & 0 \\
0 & e^{i k_0 h q_{1, 0}} & 0 & 0 \\
0 & 0 & 1 & 0 \\
0 & 0 & 0 & 1
\end{bmatrix*}, \\
\mathbf{Q}_{i,\leftarrow}
& = &
\begin{bmatrix*}[c]
1 & 0 & 0 & 0 \\
0 & 1 & 0 & 0 \\
0 & 0 & e^{i k_0 h q_{i, 2}} & 0 \\
0 & 0 & 0 & e^{i k_0 h q_{i, 3}}
\end{bmatrix*}.
\end{array}
\end{equation}
$\mathbf{E}_{i, \rightarrow} (z_{i+1})$ and $\mathbf{E}_{i, \leftarrow} (z_{i+1})$ can then be eliminated from Eq.~\eqref{eq:split_eigen},
\begin{equation}
\label{eq:kragt_equation}
\mathbf{P}_{i,\text{out}} \mathbf{Q}_{i,\rightarrow}
\begin{bmatrix*}[c]
\mathbf{E}_{i, \rightarrow} (z_{i}) \\
\mathbf{E}_{i+1, \leftarrow} (z_{i+1})
\end{bmatrix*}
=
\mathbf{P}_{i,\text{in}} \mathbf{Q}_{i,\leftarrow}
\begin{bmatrix*}[c]
\mathbf{E}_{i+1, \rightarrow} (z_{i+1}) \\
\mathbf{E}_{i, \leftarrow} (z_{i})
\end{bmatrix*}.
\end{equation}
This eventually leads to the definition of the \textit{scattering matrix} $\mathbf{S}_{i,i+1}$ between interfaces $i$ and $i+1$, as
\begin{equation}
\label{eq:scat_equation_2}
\begin{bmatrix*}[c]
\mathbf{E}_{i+1, \rightarrow} (z_{i+1}) \\
\mathbf{E}_{i, \leftarrow} (z_{i})
\end{bmatrix*}
=
\mathbf{S}_{i,i+1}
\begin{bmatrix*}[c]
\mathbf{E}_{i, \rightarrow} (z_{i}) \\
\mathbf{E}_{i+1, \leftarrow} (z_{i+1})
\end{bmatrix*}
\end{equation}
with
\begin{equation}
\label{eq:scat_equation}
\mathbf{S}_{i,i+1} = \mathbf{Q}_{i,\leftarrow}^{-1} \mathbf{P}_{i,\text{in}}^{-1} \mathbf{P}_{i,\text{out}} \mathbf{Q}_{i,\rightarrow}.
\end{equation}

Evidently, the combination of scattering matrices to describe multilayer stacks cannot be as straightforward as matrix multiplications in the case of the transfer matrix method. For example, the construction of $\mathbf{S}_{i,i+2}$ would require removing $\mathbf{E}_{i+1,\rightarrow}$ and $\mathbf{E}_{i+1,\leftarrow}$. As above, this can be done by linear operations.

Rewriting the scattering matrices in terms of their $2 \times 2$ quadrants,
\begin{equation}
\begin{array}{rcl}
\mathbf{S}_{i,i+1}
& = &
\begin{bmatrix*}[c]
\mathbf{S}^{(1)}_{00} & \mathbf{S}^{(1)}_{01} \\
\mathbf{S}^{(1)}_{10} & \mathbf{S}^{(1)}_{11}
\end{bmatrix*}, \\
\mathbf{S}_{i+1,i+2}
& = &
\begin{bmatrix*}[c]
\mathbf{S}^{(2)}_{00} & \mathbf{S}^{(2)}_{01} \\
\mathbf{S}^{(2)}_{10} & \mathbf{S}^{(2)}_{11}
\end{bmatrix*}, \\
\mathbf{S}_{i,i+2}
& = &
\begin{bmatrix*}[c]
\mathbf{S}^{(0)}_{00} & \mathbf{S}^{(0)}_{01} \\
\mathbf{S}^{(0)}_{10} & \mathbf{S}^{(0)}_{11}
\end{bmatrix*},
\end{array}
\end{equation}
eventually leads to the following expressions for the quadrants of $\mathbf{S}_{i,i+2}$,
\begin{equation}
\label{eq:combination_scatmat}
\begin{array}{lcl}
\mathbf{S}^{(0)}_{00} & = & \mathbf{S}^{(2)}_{00} \mathbf{C}^{-1} \mathbf{S}^{(1)}_{00}, \\
\mathbf{S}^{(0)}_{01} & = & \mathbf{S}^{(2)}_{01} + \mathbf{S}^{(2)}_{00} \mathbf{C}^{-1} \mathbf{S}^{(1)}_{01} \mathbf{S}^{(2)}_{11}, \\
\mathbf{S}^{(0)}_{10} & = & \mathbf{S}^{(1)}_{10} + \mathbf{S}^{(1)}_{11} \mathbf{S}^{(2)}_{10} \mathbf{C}^{-1} \mathbf{S}^{(1)}_{00}, \\
\mathbf{S}^{(0)}_{11} & = & \mathbf{S}^{(1)}_{11} \mathbf{S}^{(2)}_{11} + \mathbf{S}^{(1)}_{11} \mathbf{S}^{(2)}_{10} \mathbf{C}^{-1} \mathbf{S}^{(1)}_{01} \mathbf{S}^{(2)}_{11}, \\
\end{array}
\end{equation}
with
\begin{alignat}{2}
\mathbf{C} & = & \begin{bmatrix*}[c] 1 & 0 \\ 0 & 1 \end{bmatrix*} - \mathbf{S}^{(1)}_{01} \mathbf{S}^{(2)}_{10}
\end{alignat}

Carrying this combination for all layers of the stack, including the isotropic entry and exit media (with analytical formulas for their eigenvectors and with their propagation matrix set to identity), and expressing the forward- and backward-propagating electric fields in terms of incident, reflected and transmitted waves eventually leads to the total scattering matrix of the multilayered stack,
\begin{equation}
\label{eq:full_scattering}
\begin{bmatrix*}[c]
E_{p,\rightarrow}(z_N) \\ E_{s,\rightarrow} (z_N)  \\ E_{p,\leftarrow}(z_0) \\ E_{s,\leftarrow} (z_0) 
\end{bmatrix*}
=
\mathbf{S}
\begin{bmatrix*}[c]
E_{p,\rightarrow}(z_0)  \\ E_{s,\rightarrow}(z_0) \\ E_{p,\leftarrow}(z_N) \\ E_{s,\leftarrow}(z_N),
\end{bmatrix*}.
\end{equation}

The reflection and transmission coefficients can be then obtained straightforwardly from the scattering matrix elements,
\begin{equation}
\label{eq:reflco_sm}
\begin{array}{lclcl}
r_{\text{pp}} & = & S_{20} \\
r_{\text{ps}} & = & S_{21} \\
r_{\text{sp}} & = & S_{30} \\
r_{\text{ss}} & = & S_{31} \\
t_{\text{pp}} & = & S_{00} \\
t_{\text{ps}} & = & S_{01} \\
t_{\text{sp}} & = & S_{10} \\
t_{\text{ss}} & = & S_{11}
\end{array}
\end{equation}

The reflection and transmission coefficients in the circular polarisation basis can finally be obtained from Eqs.~\eqref{eq:jones_circ_1} and \eqref{eq:jones_circ_2}, respectively.

\section{Implementation}
\label{sec:code_structure}

\subsection{Code architecture} 
\label{sec:code_architecture}

\begin{figure}
  \centering
    \includegraphics[width=\textwidth]{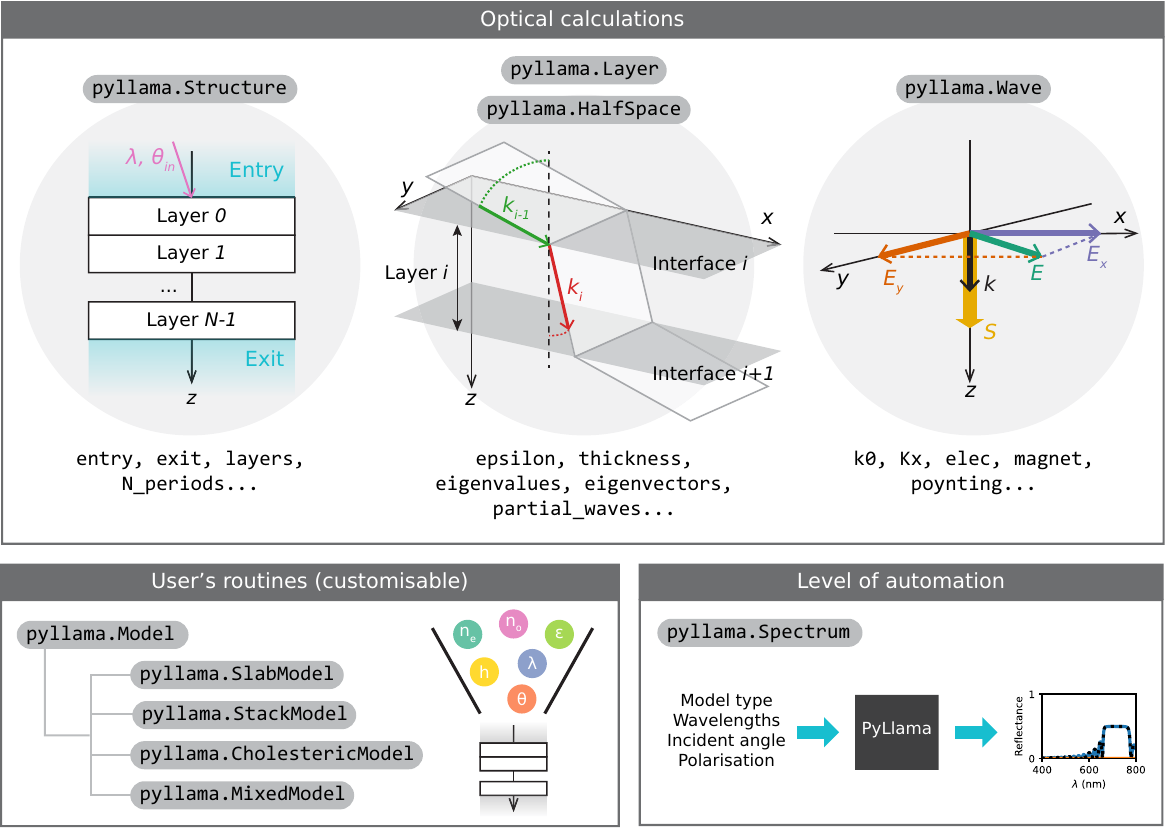}
	\caption{Schematic of the code organisation. The optical calculations are carried out in the classes \texttt{Structure}, \texttt{Layer} with its subclass \texttt{HalfSpace} and \texttt{Wave}. The children classes of \texttt{Model} generate \texttt{Structures} from chosen parameters according to the user's need and are more convenient to handle. Some models are predefined (a slab with \texttt{SlabModel}, a periodic stack with \texttt{StackModel}, a cholesteric with \texttt{CholestericModel} and a combination of several models with \texttt{MixedModels}) and the user may write their own additional routines. Lastly, the class \texttt{Spectrum} give an additional level of automation and allows the user to use the code as a black box, when and if needed.}
	\label{fig:code_structure}
\end{figure}

PyLlama is practically structured to separate the implementation of the theory from the user, and at the same time to allow the user to access details about the optical calculations on demand. As schematised on Fig.~\ref{fig:code_structure}, the code is organised into:
\begin{itemize}
    \item Classes that carry the optical calculations: the class \texttt{Structure} represents the multilayer stack and lists the layers it consists in; the class \texttt{Layer} represents one layer and its partial waves, its child class \texttt{HalfSpace} being the equivalent for the entry and exit semi-infinite media; and the class \texttt{Wave} represents optical waves.
    
    \item Classes that contain routines to construct \texttt{Structures} from appropriate parameters. They are children of a master class \texttt{Model} and solely serve the purpose of building \texttt{Structures} in a way that is easier to manipulate for the user than the class \texttt{Structure} itself. A few models are already implemented in PyLlama.
    
    \item A class \texttt{Spectrum} that allows for a further level of automation for common operations such as calculating a full spectrum (in linear or circular polarisation) and exporting the calculated data. This class enables the use of the code as a black-box -- though any detail on the optics remains accessible.
\end{itemize}

\subsection{Optical calculations}
\label{sec:optical_calculations}

Each partial wave in PyLlama is represented by the class \texttt{Wave} that contains fields for the electric field (\texttt{Wave.elec}), the magnetic field (\texttt{Wave.magnet}), the Poynting vector (\texttt{Wave.poynting}) as well as the constant component of the wavevector $K_x = q_j$ (\texttt{Wave.Kx}).

The class \texttt{Layer} contains fields for the layer permittivity $\epsilon$ (\texttt{Layer.eps}), its thickness $h$ (\texttt{Layer.thickness}), and the wavevector constant projection on the $x$~axis $k_x = K_x k_0$ (\texttt{Layer.Kx} and \texttt{Layer.k0}). With these parameters, the $4\times 4$ matrix $\bm{\Delta}$ (\texttt{Layer.D}) is calculated with the function \texttt{Layer.build\_D()} from the other fields of \texttt{Layer}. The class \texttt{Layer} also implements the calculation of the partial waves described in Section~\ref{sec:one_layer} as a list of four \texttt{Wave} instances stored in a field \texttt{Layer.partial\_waves}. The entry and exit isotropic media are represented by the class \texttt{HalfSpace}, which is a children of \texttt{Layer} with the only differences residing in the calculation of the eigenvectors (calculated analytically) and eigenvalues (set to $1$ for no propagation phenomena).

The eigenvalues and eigenvectors are calculated from the matrix $\bm{\Delta}$ (\texttt{Layer.D}) with Python’s numerical package Numpy or Python’s symbolic package Sympy. The eigenvalues are unique but the eigenvectors are defined up to a normalisation factor that depends on the algorithm (for example, Numpy normalises each eigenvector so that its module is 1). The eigenvectors calculated with one or another numerical method might therefore differ by a factor $-1$ or $i$, which will inhibit their sorting in the way that is described in~\ref{sec:partial_waves}. This does not prevent the use of the transfer or scattering matrix methods to go from one layer to the next. However, the eigenvectors of the entry and exit \texttt{HalfSpaces} need to be set in a specific order to allow for the decomposition of the incident, reflected and transmitted electric fields into their appropriate components. In general, if the eigenvectors are calculated analytically (formulas may be found in the literature for specific anisotropic media~\cite{Yeh1988, Palto2001, Eidner1989, Kragt2019, Oldano1989}), the analytical formulas enable to sort them as forward and backward pairs with distinct polarisations (as p and s in the isotropic entry and exit semi-infinite spaces) and this would allow the user to access intermediate transfer or scattering matrices for specific layers in the stack.

One key point emerging from the $4 \times 4$ matrix formalism is the possibility to treat all layers of the stack independently: once the matrix $\bm{\Delta}$ has been calculated for each layer (and the eigenvalues, eigenvectors and partial waves that arise from it) as instances of the \texttt{Layer} class, the \texttt{Layers} can be treated as building blocks to build a stack. In PyLlama, we represent the multilayer stack as a \texttt{Structure}, which contains a list of \texttt{Layers} (field \texttt{Structure.layers}) as well as the entry and exit half-spaces (fields \texttt{Structure.entry} and \texttt{Structure.exit}). Once all \texttt{Layers} of the \texttt{Structure} have been characterised, they can be combined together with the half-spaces, either with the transfer matrix method or with the scattering matrix method, to obtain the reflection and transmission coefficients.

The class \texttt{Structure} contains several functions that implement the optical calculations:
\begin{itemize}
\item \texttt{Structure.build\_transfer\_matrix()} calculates the transfer matrix of the stack (entry half-space, consecutive layers, and exit half-space)
\item \texttt{Structure.\_build\_scattering\_matrix\_to\_next(this\_layer, next\_layer)} calculates the scattering matrix for the layer \texttt{this\_layer} and the interface between \texttt{this\_layer} and the following layer \texttt{next\_layer}
\item \texttt{Structure.\_combine\_scattering\_matrices(S\_ab, S\_bc)} combines the scattering matrices $\mathbf{S}_{a,b}$ \texttt{S\_ab} and $\mathbf{S}_{b,c}$ \texttt{S\_bc} into the scattering matrix $\mathbf{S}_{a,c}$
\item  \texttt{Structure.build\_scattering\_matrix()} calculates the scattering matrix for the complete multilayer stack
\item \texttt{Structure.get\_fresnel(method=<"SM"|"TM">)} calculate the reflection and transmission coefficients, which take the form of a Jones matrix. The user may choose the underlying matrix method with the argument \texttt{method}. The reflection coefficients can be converted to circular polarisation with the function \texttt{Structure.fresnel\_to\_fresnel\_circ}.
\item \texttt{Structure.get\_refl\_trans(method=<"SM"|"TM"|"EM">, circ=<True|False>)} calculates the reflectance and transmittance directly. The argument \texttt{circ} enables to choose between linear and circular polarisation bases. The transfer matrix can also be calculated with the direct exponential of $\bm{\Delta}$ (without the diagonalisation) through the choice \texttt{"EM"}.
\end{itemize}

The user manual contains tutorials and a detailed documentation containing all implemented classes and functions as well as their parameters and returns.

\subsection{Periodic stacks}
\label{sec:periodic_stacks}

Many interesting multilayer stacks are periodic (such as Bragg stacks and cholesterics). Their optical properties can of course be modelled by building a periodic structure ``by hand'' (e.g., a periodic list of permittivities) but they can also be characterised by their repeating unit to speed up the calculations: transfer matrices and scattering matrices can therefore be calculated for a repeating unit only, and combined together in a second step.

The basic situation is straightforward for the transfer matrix method. As shown in Eq.~\eqref{eq:berprod}, the propagator $\mathbf{R}$ of a stack of $N$~layers (excluding the entry and exit isotropic half-spaces) is equal to the product of the propagators of all layers, $\mathbf{R} = \overset{\curvearrowleft}{\prod}_{i=0}^{N-1} \mathbf{R}_i$. When the $N$~layers of the stack correspond to $N_\text{per}$~repeating units made of $n$~layers each, the propagator of the repeating unit is $\mathbf{R}_\text{per} = \overset{\curvearrowleft}{\prod}_{i=0}^{N-1} \mathbf{R}_i$ such that the transfer matrix of the $N_\text{per}$ repeating units is $\mathbf{R} = \overset{\curvearrowleft}{\prod}_{i=0}^{N_\text{per}-1} \mathbf{R}_\text{per}$.

To speed up the computation time, the decomposition of the stack is different than multiplying one matrix per repeating unit. The total number of repeating units $N_\text{per}$ is decomposed in powers of two: $N_\text{per} = \sum 2^k$ for the appropriate list of $k_p$s (the list of $k_p$s associated with $N_\text{per}$ is the binary form of $N_\text{per}$: we obtain it by converting $N_\text{per}$ to binary). We calculate the transfer matrix for sub-units that contain $2^{m}$ repeated units $\mathbf{R}_\text{per}^{m}$ for $m$ between $0$ and the higher power of two in the decomposition, and store them in a database. Then, we select the matrices that we need to build up the full system of $N_\text{per}$ repeated units.

For the scattering method, Eq.~\eqref{eq:combination_scatmat} shows how to combine $\mathbf{S}_{i,i+1}$ (the scattering matrix of the layer $i$ and its next interface with the layer $i+1$)  with $\mathbf{S}_{i+1,i+2}$ (the scattering matrix of the layer $i+1$ and its next interface with the layer $i+2$) into the scattering matrix $\mathbf{S}_{i,i+2}$, which encompasses the layer $i$, the layer $i+1$, and the interface from the layer $i+1$ to the layer $i+2$. For a periodic stack, the repeating unit starts at the layer $0$ and finishes at the interface between the layer $N-1$ layer and the layer $0$ of the next repeating unit. However, the last period of the stack does not end by an interface between the layer $N-1$ and the layer $0$, but with an interface between the layer $N-1$ and the isotropic half-space. Therefore, the combination of the repeated motive is only done $N_\text{per}-1$ times (with the decomposition of $N_\text{per} - 1$ into powers of two) for the $N_\text{per} - 1$ first periods, and the last period is added afterwards.

\subsection{Useful routines and customisation}
\label{sec:model}

In principle, a direct implementation of the theory behind the transfer matrix or the scattering matrix methods is sufficient to obtain the reflectance and transmittance for a multilayer stack once the layers have been properly constructed. However, the construction of the layers from scratch can become tedious. We designed the class \texttt{Model} and its children as user-friendly routines that build \texttt{Structures} made of appropriate \texttt{Layers} and \texttt{HalfSpaces}. \texttt{Model} is a parent class in which we set features that are common to all models and we wrote a few children classes,  \texttt{SlabModel}, \texttt{StackModel}, \texttt{StackOpticalThicknessModel} and \texttt{CholestericModel}. The user can therefore write a custom layer-building block of code, without having to re-write any of the underlying theory, with suitable parameters that are relevant to the physical stack being modelled.

A \texttt{Model} essentially represents an empty multilayer stack, consisting in two semi-infinite media separated by no layer. The idea behind the code architecture is that the children classes of \texttt{Model} contain additional parameters used to build a \texttt{Structure} that is relevant to the stack, for example:
\begin{itemize}
    \item \texttt{SlabModel} represents a slab of homogeneous material and its parameters are its permittivity (optionally calculated from an inputted permittivity and a rotation angle (in radians) \texttt{rotangle\_rad} around an arbitrary axis \texttt{rotaxis}) \texttt{eps}, its thickness (in nanometers) \texttt{thickness}.
    
    \item \texttt{StackModel} represents an arbitrary multilayer stack and its parameters are a list of permittivities \texttt{eps\_list} and a list of thicknesses \texttt{thickness\_list} corresponding to its layers, as well as a number of periods \texttt{N\_per} that could in principle be set to one if the lists of permittivities and thicknesses contain an item for each layer.
    
    \item \texttt{StackOpticalThicknessModel} represents a stack of isotropic layers which all have the same optical thickness and its parameters are a list of refractive indices \texttt{n\_list} and the total thickness of the stack \texttt{total\_thickness}.
    
    \item \texttt{CholestericModel} represents a cholesteric material made of birefringent units and its parameters are a \texttt{Cholesteric} object representing the (potentially distorted) helicoidal architecture \texttt{cholesteric}, the extraordinary and ordinary refractive indices of the birefringent units \texttt{n\_e} and \texttt{n\_o} and the number of repeating units where one unit is \texttt{cholesteric}. Our \texttt{CholestericModel} class interacts with our \texttt{Cholesteric} library: a \texttt{Cholesteric} instance is a parameter to build a \texttt{CholestericModel}.
\end{itemize}

Each children class of \texttt{Model} contains a redefinition of the parent function \texttt{Model.build\_structure()} that overrides it and creates \texttt{Layers} in a routine that uses the model parameters. The user manual contains tutorials explaining how to create custom children classes appropriately.

\subsection{Practical examples}
\label{sec:examples}

\begin{figure}
  \centering
    \includegraphics[width=\textwidth]{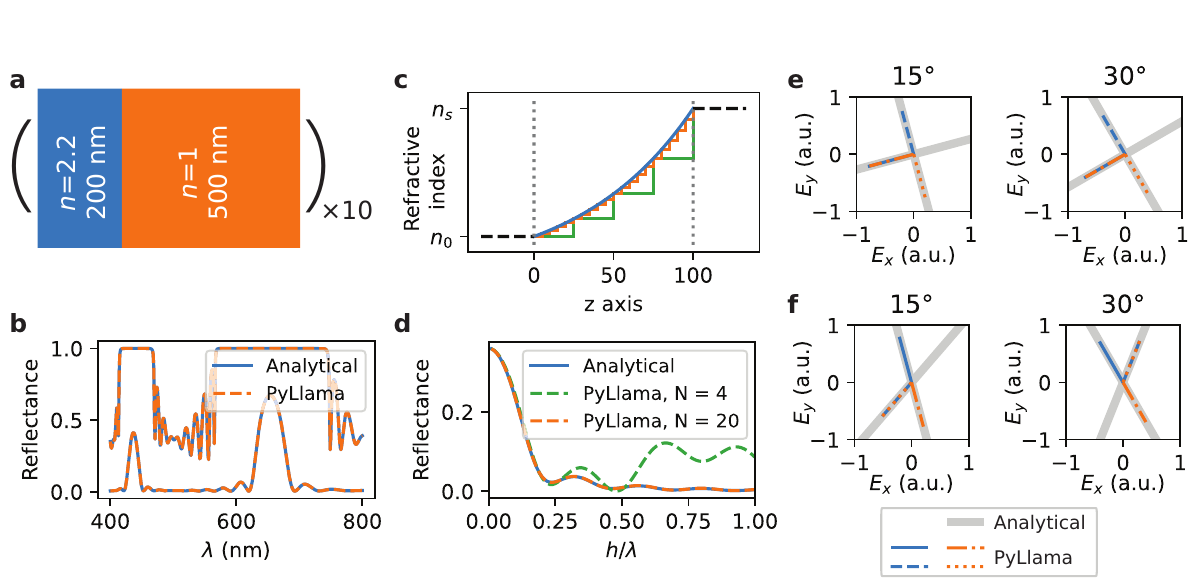}
	\caption{Comparison of PyLlama's results with other methods in the case of isotropic layers and uniaxial crystals. a, b) Isotropic Bragg stack. a) Schematic of the situation: the system consists in a periodic pattern of two layers that are repeated 10~times (the entry and exit half-spaces are the same as the first and last layers). b) Reflection spectrum of the isotropic Bragg stack calculated with PyLlama and with Yeh's matrix method~\cite{Yeh1988} for an angle of incidence $\theta_{\text{in}} = \ang{60}$. c, d) Isotropic non-homogeneous slab with exponentially varying refractive index. c) Refractive index profile: index varying continuously (blue), discretely with 20~layers of the same thickness (orange) and discretely with 4~layers of the same thickness (green). d) Reflectance calculated with PyLlama in the discrete scenarios and with Yeh's analytical formulas in the continuous scenario~\cite{Yeh1988}. e, f) Uniaxial crystals extracted from a cholesteric structure with a rotation angle around the $z$ axis $\phi_i = \ang{15}$ and $\ang{30}$. The direction of the electric field of the four partial waves calculated numerically with PyLlama matches these calculated analytically~\cite{Oldano1989}. The angle of incidence is e) $\ang{0}$ and f) $\ang{60}$.}
	\label{fig:validation_isotropic}
\end{figure}

Figure~\ref{fig:validation_isotropic} shows examples of results obtained with our different models and validated with available analytical formulas. Figure~\ref{fig:validation_isotropic}a and b display the case of an isotropic Bragg stack. The system consists in a periodic unit of two layers that are repeated 10~times, including the entry and exit half-spaces that are the same as the first and last layers (with no thickness). The reflection spectra of the isotropic Bragg stack calculated with our code and the model \texttt{StackModel} agrees with Yeh's matrix method~\cite{Yeh1988} for s and p-polarisations. Here we display spectra for an angle of incidence $\theta_{\text{in}} = \ang{60}$.

Figure~\ref{fig:validation_isotropic}c and d display the case of an isotropic non-homogeneous slab analysed by Yeh~\cite{Yeh1988}. The slab has a thickness $L$ in the order of hundreds of nanometers, a refractive index $n_0$ at $z=0$ and a refractive index $z_s$ at $z=L$. The refractive index profile across the slab is continuous with an exponential profile defined by:
\begin{equation}
\label{eq:yeh_exponential}
    n(z) = n_0 \left( \frac{n_s}{n_0} \right) ^{z/L}
\end{equation}
Yeh gives analytical formulas for the reflection spectrum and also shows analytically that a series of $N$ discrete layers with appropriate thicknesses and refractive indices can be used to model the non-homogeneous slabs, with a precision that increases when the resolution $N$ increases. We used our model \texttt{StackModel} and input a list of $N$ permittivities (calculated from the $N$ refractive indices) and a list of $N$ thicknesses, all identical. We could also have used \texttt{StackModel} and calculated the individual thicknesses outside, it was more convenient to create a dedicated model. If we were to work often with non-homogeneous slabs, we could also have imagined a model \texttt{NonHomogeneousSlabModel} taking a function $n(z)$ (\texttt{lambda z : n0 * (ns / n0) ** (z / L)} in our case) as a parameter and automatically handling the splitting in layers of the same thickness. This illustrate the main idea behind the creation of models: the user can write up the classes (children of \texttt{Model}) that they wish in order to model structures by handling the parameters that the user think are convenient.

Figure~\ref{fig:validation_isotropic}e shows the direction of the eigenvectors computed for a single layer of nematic crystal (taken here as an individual \texttt{Layer} in a \texttt{Structure} generated by a \texttt{CholestericModel}) and compares it with analytical predictions~\cite{Oldano1989}. The agreement is quantitative. It is interesting to note that the electric field follows the cholesteric rotating director at normal incidence, as predicted by De Vries~\cite{DeVries1951} and confirmed here.

Cholesterics are of special concern in the present work. For modelling purposes, these continuous non-homogeneous materials are assimilated to discrete stacks of slices where each slice $i$ (at depth $s_i$ on the helical axis) possesses a director $n_i$ and rotation angle $\phi_i$, leading to a permittivity tensor $\epsilon(\phi_i)$. A more general discussion on cholesterics is provided in \ref{sec:cholesteric}.

\begin{figure}
  \centering
    \includegraphics[width=\textwidth]{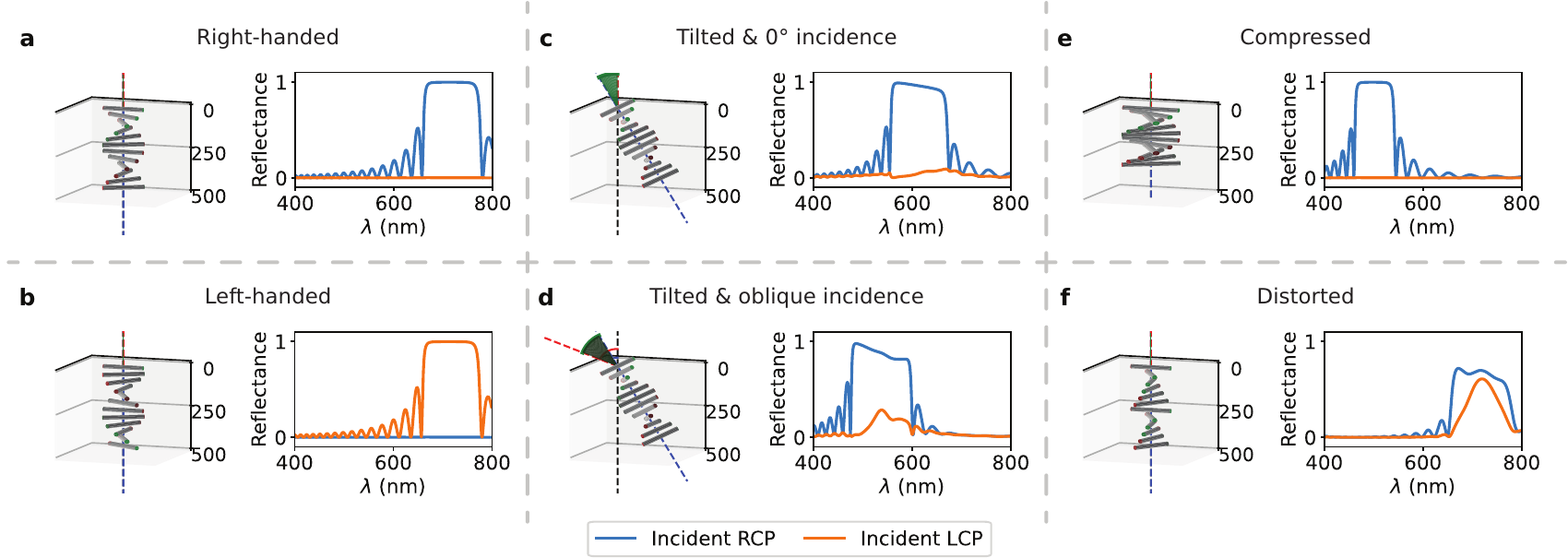}
	\caption{Collection of cholesteric-specific situations implemented in PyLlama (right-handed cholesteric, left-handed cholesteric, tilt, oblique incidence, vertical compression, distortion combining a tilt and a vertical compression), and their reflection spectra spectra calculated numerically. The user can easily mix-and-match these situations. Both the 3D representations of the cholesterics and the optical spectra are automatically computed through PyLlama.}
	\label{fig:showcase}
\end{figure}

The \texttt{Cholesteric} library allows the user to manipulate (distort, compress) the structure in a physical way, and then export the slices’ directors to build up the corresponding multilayer stack from which to calculate the reflectance. Details about the physical model’s derivation and how it handles vertical compression may be found in Ref.~\cite{Frka-Petesic2019}. This enables us to easily model different configurations of practical interest. Figure~\ref{fig:showcase} showcases a few configurations that our \texttt{Cholesteric} library can generate and that we can optically model: a right-handed cholesteric, a left-handed cholesteric, oblique incidence of light upon a straight or tilted cholesteric, a vertically compressed cholesteric, and a distorted cholesteric. To account for the tilt of a helicoid, we set the $z$-axis along the helical axis and we calculate the effective angle of incidence upon the stack with the absolute angle of incidence and the tilt of the \texttt{Cholesteric}.

As we can see from the cases represented here, the handedness, the angle of incidence and the distortion strongly impact the peak wavelength, the peak shape and the polarisation selectivity (reflectance from unpolarised light can even be higher than 50\% in some cases). Each time, a \texttt{Cholesteric} object describes the physical architecture and the required parameters are extracted to build up an optical model upon it and enables to calculate reflection spectra in the circular-polarisation basis.

We use this example to illustrate results obtained with PyLlama and to demonstrate how flexible it is to handle custom structures. The user may add supplementary features to our \texttt{Cholesteric} library, or pair an external library to PyLlama in a similar way to model custom stacks.

\subsection{Mixing multiple stacks}
\label{sec:mixed_model}

\begin{figure}
  \centering
    \includegraphics[width=\textwidth]{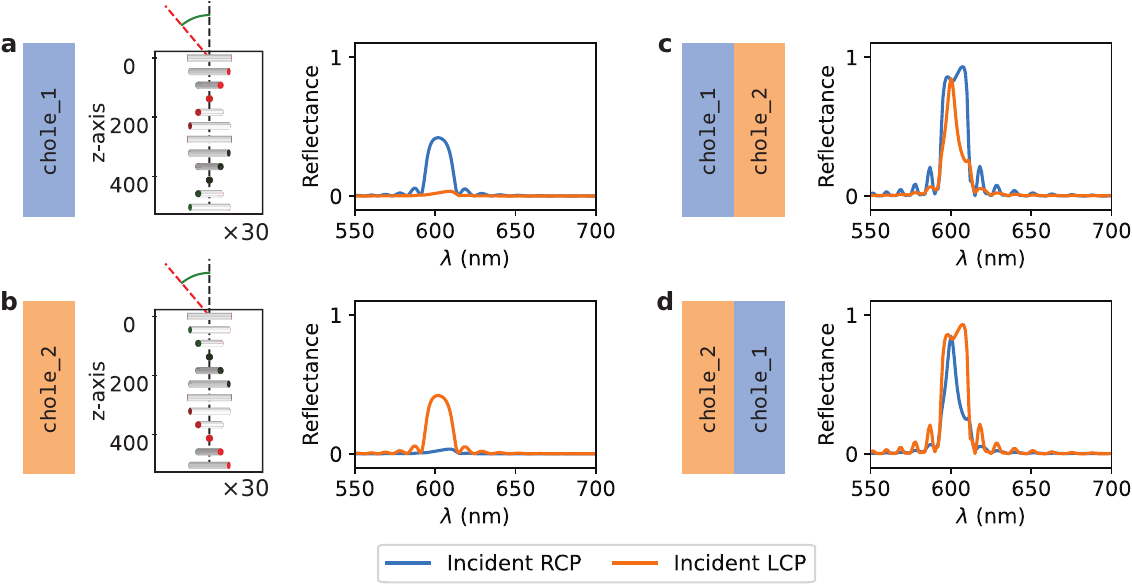}
	\caption{a, b) Schematic representation and reflection spectra under $\ang{45}$ incidence of a) a right-handed and b) a left-handed cholesteric with the same pitch (500~nm) and with the same number of pitches (30). c, d) Reflection spectra for the stacks of c) the right-handed cholesteric on top of the left-handed cholesteric and d) the left-handed cholesteric on top of the right-handed cholesteric.}
	\label{fig:mixed_model}
\end{figure}

Throughout our code, \texttt{Layers} are treated as independent units and can be added to a \texttt{Structure} provided that the wavevector $x$-component stays the same. Similarly, \texttt{Structures} can be stacked together to model the behaviour of complex samples such as understanding the self-assembly of dry cellulose nanocrystal films that creates distorted domains~\cite{Frka-Petesic2020}, explaining the polarization independence of hydroxypropyl cellulose films distorted by cross-linking~\cite{Chan2019} or tuning the distortion to obtain samples with a particular polarization selectivity~\cite{Kragt2019} or gaining insight on mechanical deformations imposed on cellulose nanocrystal elastomers~\cite{Boott2020}. Combining models into a master model allows to take advantage of sub-periodicities within each model and to keep the computation time fast.

The class \texttt{MixedModel}, child of \texttt{Model}, takes as additional parameter a list of models \texttt{models\_list} to be combined (in the order of the list). They are screened in order to add only the models that are compatible with the first one from the list (in terms of the conservation of the wavevector’s tangential component). The function \texttt{MixedModel.get\_refl()} overrides its parent function \texttt{Model.get\_refl()} and calls for \texttt{get\_refl\_TM\_multiple\_structures} (respectively, \texttt{SM}), which builds up transfer or scattering matrices for each sub-model and combines them with a cautious handling the interfaces between consecutive models for the scattering matrix method.

An example is given on Fig.~\ref{fig:mixed_model} where two periodic \texttt{CholestericModels} of opposite handedness and same pitch are stacked upon each other in a different order. The resulting spectrum for the stacked cholesterics under oblique incidence show a difference depending on which cholesteric is on top of the other.

\subsection{A level of automation}
\label{sec:spectrum}

The classes \texttt{Structure}, \texttt{Layer}, \texttt{HalfSpace} and \texttt{Wave} contain every function that is necessary to calculate the reflectance of a mutlilayer stack at a specific wavelength, and the class \texttt{Model} and its children enable to build \texttt{Structures} easily from appropriate parameters.

In order to enable the use of our code without requiring extensive programming skills and/or manipulation of the theoretical concepts to calculate reflection spectra for a given multilayer stack, we embedded our models inside the class \texttt{Spectrum} that provides a level of automation to the user to calculate the reflectance and transmittance for a range of wavelengths (to get a full spectrum) and export the data for further analysis in Python or in MATLAB, while the optics calculations are occurring in the background. The user who decides to write custom models can integrate them to the class \texttt{Spectrum} too.

\texttt{Spectrum} contains a list of wavelengths inputted by the user, a dictionary whose keys correspond to the model parameters and whose values correspond to the values assigned to these parameters inputted by the user, and an empty dictionary which will contain the calculated reflection spectra. The function \texttt{Spectrum.calculate\_refl\_trans()} enables to calculate reflection spectra in the linear polarisation basis or in the circular polarisation basis with a choice of the matrix method with the parameter, and with a display of the calculation progress with the parameter \texttt{talk=<False|True>}. Results for (potentially both) polarisation bases are then added to \texttt{Spectrum.data}. The function \texttt{Spectrum.export()} enables to export the content of \texttt{Spectrum.data} for further processing in Python or in MATLAB. Its argument is a filename \texttt{path\_out} which extension should be \texttt{.pck} for an export with Pickles and further use in Python or \texttt{.mat} for an export in MATLAB-compatible format.

When pairing \texttt{Model} and its children with the class \texttt{Spectrum}, spectra from complex structures are straightforward to obtain even without much experience in programming. The custom \texttt{Models} created by the user can also be with the class \texttt{Spectrum}. The user manual contains tutorials explaining how to incorporate custom children classes appropriately.

\section{Comparison between the transfer matrix and scattering matrix methods}
\label{sec:comp_TM_SM}

The calculations behind the transfer matrix and the scattering matrix methods have been detailed in Section~\ref{sec:transfer_matrix} and~\ref{sec:scattering_matrix}, and show that the combination of subsequent \texttt{Layers} with the transfer matrix method simply consist in matrix multiplications, Eq.~\eqref{eq:berprod} while their combination with the scattering matrix method requires to unpack matrix elements, to combine them externally and to reassemble them into a new scattering matrix, Eq.~\eqref{eq:combination_scatmat}. The scattering matrix method is consequently slower than the transfer matrix method, but it is also more robust and stable.

\begin{figure}
  \centering
    \includegraphics[width=\textwidth]{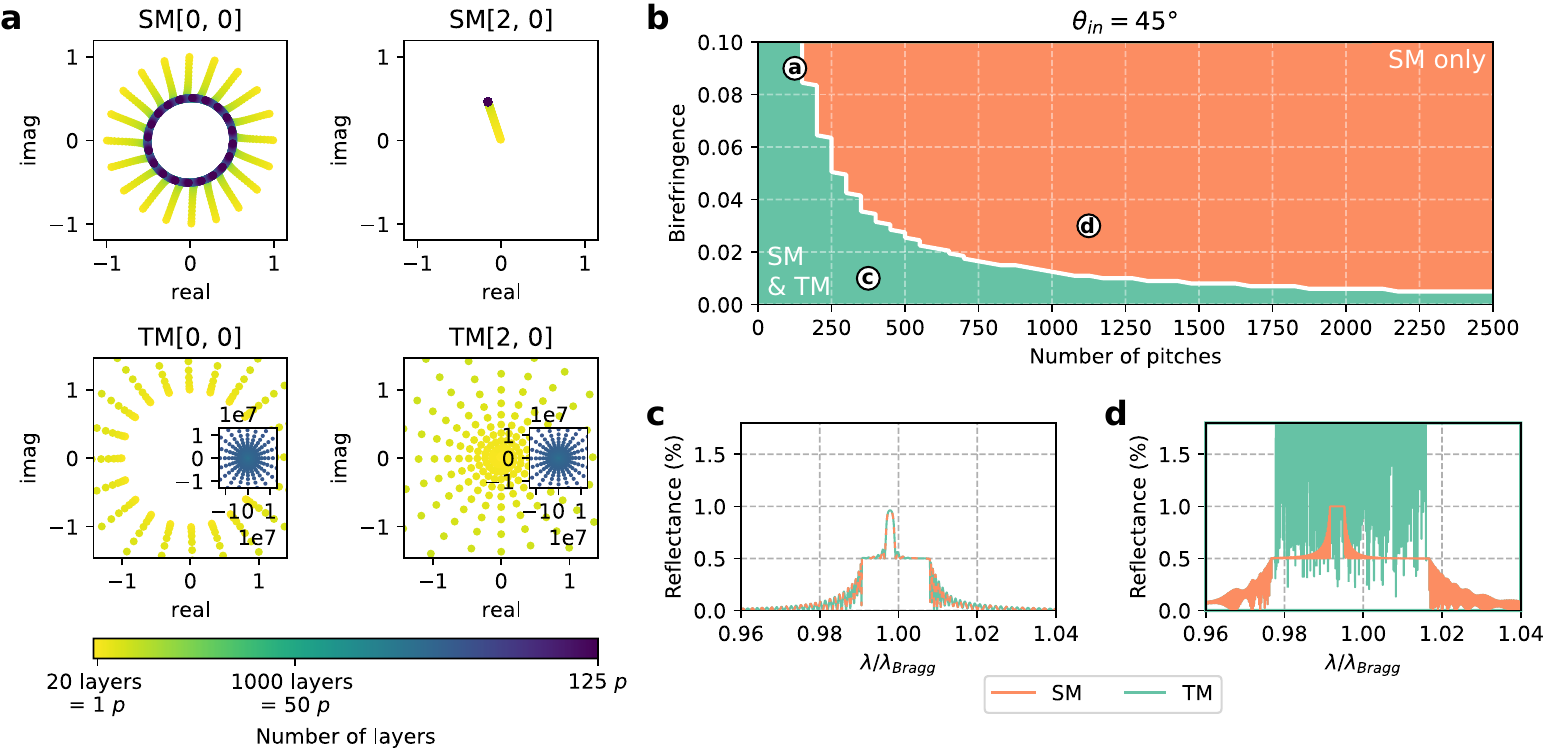}
	\caption{Comparison between the scattering matrix (SM) and transfer matrix (TM) methods a) Evolution of two matrix elements of indices $(0, 0)$ and $(2, 0)$ in the complex plane. The size of the system increases from one layer to a full pitch made of 20 layers to 125~pitches made of $125\times20$ layers. b) The colormap displays the combinations that cannot be calculated with the TM method. The configurations a), b) and c) are pinpointed on the colormap. c, d) The TM and SM methods give identical results at low birefringence and for thin systems (c) while the TM method becomes numerically unstable at high birefringence and for thick systems (d). The pitch and the Bragg wavelength $\lambda_{Bragg}$ are given in~\ref{sec:cholesteric}: the pitch $p$ is the periodicity and the Bragg wavelength equals $n p \cos(\theta_{in})$, where $n$ is the average refractive index of the cholesteric.}
	\label{fig:performances}
\end{figure}

Figure~\ref{fig:performances}a shows the evolution of selected (representative) matrix elements from the transfer and scattering matrices when the number of layers in a cholesteric structure increases (Appendix~\ref{sec:cholesteric} provides details about the structure). They are plotted in the complex plane. As one can see, the values obtained with the transfer matrix method are continuously growing, while the values obtained with the scattering matrix method are converging. In fact, at each calculation step, the values of the scattering matrix represent transmission and reflection coefficients, Eq.~\eqref{eq:reflco_sm}, and their modulus is therefore always under 1, while with the transfer matrix method, reflection coefficients are calculated as ratios between given matrix elements, Eq.~\eqref{eq:reflco_tm}.

Usually, reflection spectra calculated from both methods lead to the same result (Figure~\ref{fig:performances}c: a cholesteric structure with a birefringence $n_e - n_o = 0.015$ and  375 pitches). However, when the number of layers becomes too large or the birefringence becomes too high, the transfer matrix method fails dramatically (Figure~\ref{fig:performances}d: a cholesteric structure with $n_e - n_o = 0.035$ and  1125 pitches) while the scattering matrix remains stable. Figure~\ref{fig:performances}b displays which combinations of birefringences and number of pitches are not compatible with the transfer matrix method.

The transfer matrix method remains useful as it is faster than the scattering matrix method: computing a spectrum on the visible range (400 to 800~nm with a resolution of 1~nm) corresponding to the cholesteric structure on Figure~\ref{fig:performances}c takes on average 6.5s with the transfer matrix method against 14.0s with the scattering matrix method (averaged over 10 runs with HP EliteBoox x360 1030 G2, with a standard deviation under 0.1).

\section{Concluding remarks}
\label{sec:concluding_remarks}

We have presented a stable and versatile Python toolkit for the electromagnetic modelling of anisotropic multilayer stacks. PyLlama relies on known concepts and numerical techniques, but it answers a need from the scientific community for a simple, freely-available and open-source program for the purpose. It is accessible to use without extended programming experience even for complex multilayer stacks (distorted cholesterics, stacks containing periodic sub-structures...), and it is also easy to customise without having to re-write any optical calculation.

The strength of PyLlama is undoubtedly its capability to deal with arbitrary multilayers even in extreme situations (large number of layers, high birefringence, grazing angles of incidence) thanks to the implementation of the scattering matrix method, and to switch seamlessly to the transfer matrix method for faster computation speed, when suitable. PyLlama thus allows performing a side-by-side comparison between the transfer matrix and the scattering matrix methods.

Many additional functionalities could be added in the near future without considerable efforts, like the possibility to incorporate magneto-optic effects~\cite{schubert1999explicit}, and to compute spatial field distribution in the stack~\cite{Passler2017} and Bloch modes for periodic media~\cite{cao2002stable}. Classes dedicated to the acquisition of datasets other than reflection and transmission spectra could be created quite easily, such as for ellipsometry measurements~\cite{schubert2004infrared} or critical angle measurements~\cite{Riviere1978}, which are useful tools to characterise samples. The necessary calculations are already carried out in the core of PyLlama, which paves the way to the implementation of a dedicated framework similar to our class \texttt{Spectrum}.

The electromagnetic modelling of space-time-modulated media -- a rapidly growing topic in wave physics~\cite{galiffi2021photonics} -- would be an exciting future development of PyLlama. The transfer matrix method has recently been generalized to consider time-varying systems~\cite{li2019transfer} by writing the fields as a sum of harmonics of the modulation frequency and setting up time-varying boundary conditions. This results in matrices of larger sizes depending on the number of harmonics considered. The implementation of this idea for anisotropic materials may be of significant help to explore the multitude of exotic phenomena in such systems.

It is our hope that the scientific community will adopt PyLlama and contribute to its evolution.

\section{Acknowledgements}

This work was supported by ERC grant ERC-2014-STG H2020 639088 and Philip Leverhulme Prize (PLP-2019-271) for S.V. and M. M. B. The authors are indebted to Jean-Paul Hugonin (Laboratoire Charles Fabry, Palaiseau, France) for his critical reading of the manuscript, for verifying our numerical simulations with the RCWA software RETICOLO~\cite{hugonin2021reticolo}, and more generaly for very insightful discussions. The authors thank Bruno Frka-Petesic (Department of Chemistry, University of Cambridge, UK) for useful discussions about the optical properties of cholesteric structures. M.M.B. thanks Sing-Teng Chua (Department of Chemistry, University of Cambridge, UK) for proof-testing the user manual, and Mathieu Br\`{e}thes for his help regarding the documentation. K.V. thanks Akhlesh Lakhtakia (Penn State University, USA) for pointing out the seminal contribution of Jean Billard to the $4 \times 4$ matrix formalism~\cite{Billard1966}.

\appendix

\section{Analysis of the partial waves}
\label{sec:partial_waves}

The four partial waves in a layer can generally be separated into a forward ($+z$) propagating pair and a backward ($-z$) propagating pair, where each pair consists in a partial wave rather polarised along the $x$-axis and a partial wave rather polarised along the $y$-axis~\cite{Yeh1979}.

The main scope of the partial waves analysis is to identify the exponentially decaying and growing waves and sort them accordingly, which is important for the scattering matrix method. This can be done easily from the imaginary component of the eigenvalues (or wavevector) $q_j \equiv K_{j,z}$.

Here, we chose to rely instead on the Poynting vector along the propagation direction $z$, $\mathcal{S}_{j,z}$ (we discard here the layer label $i$), defined as
\begin{equation}
 \bm{\mathcal{S}_j} = \begin{bmatrix*} \mathcal{S}_{j,x} \\ \mathcal{S}_{j,y} \\ \mathcal{S}_{j,z} \end{bmatrix*} = \begin{bmatrix*}E_{j,y} H_{j,z} - E_{j,z}H_{j,y} \\ E_{j,z}H_{j,x} - E_{j,x}H_{j,z} \\ E_{j,x}H_{j,y} - E_{j,y}H_{j,x} \end{bmatrix*} \\
\end{equation}
with $E_{j,z}$ and $H_{j,z}$ expressed in function of $E_{j,x}$, $E_{j,y}$, $H_{j,x}$ and $H_{j,y}$,
\begin{equation}
\begin{split}
 E_{j,z}     & =  -\frac{\epsilon_{zx}}{\epsilon_{zz}}E_{j,x} -\frac{\epsilon_{zy}}{\epsilon_{zz}} E_{j,y} -\frac{K_{j,x}}{\epsilon_{zz}} H_{j,y} \\
 H_{j,z} & = K_{j,x} E_{j,y}
\end{split}
\end{equation}
The direction of the partial wave is then given by the sign of the real part of $\mathcal{S}_{j,z}$, when real, and by the sign of the imaginary part of $\mathcal{S}_{j,z}$, when complex. For the sorting of the exponentially decaying and growing waves, we have observed that this procedure was fully equivalent to that with $K_{j,z}$~\footnote{The sorting with $K_{j,z}$ is implemented and commented in the PyLlama code for the interested user.}.

An inspection of the two waves within each pair can reveal the partial wave polarisation. If the material is anisotropic (without the crystal axes along the $(x, y)$ axes), each partial wave $\mathbf{p}_j$ is analysed by calculating the ratio $C(\mathbf{p}_j)$:
\begin{equation}
\label{eq:c_poynting}
C(\mathbf{p}_j) = \frac{|\mathcal{S}_{j, x}|^2}{|\mathcal{S}_{j, x}|^2 + |\mathcal{S}_{j, y}|^2}
\end{equation}
and comparing it to its pair. If $C(\mathbf{p}_1) > C(\mathbf{p}_2)$, $\mathbf{p}_1$ describes a wave rather polarised along the $x$~axis and $\mathbf{p}_2$ describes a wave rather polarised along the $y$~axis. If the material is isotropic or anisotropic with the crystal axes aligned with the laboratory axes, this corresponds to p- and s-polarised waves, respectively. According to Ref.~\cite{Passler2017}, in these cases, the two partial waves within each pair can be analysed with their electric field $\mathbf{E}_j$ instead of their Poynting vector $\bm{\mathcal{S}}_j$:
\begin{equation}
\label{eq:c_elec}
C_\text{iso}(\mathbf{p}_j) = \frac{|E_{j,x}|^2}{|E_{j,x}|^2 + |E_{j,y}|^2}
\end{equation}

\begin{figure}
  \centering
    \includegraphics[width=\textwidth]{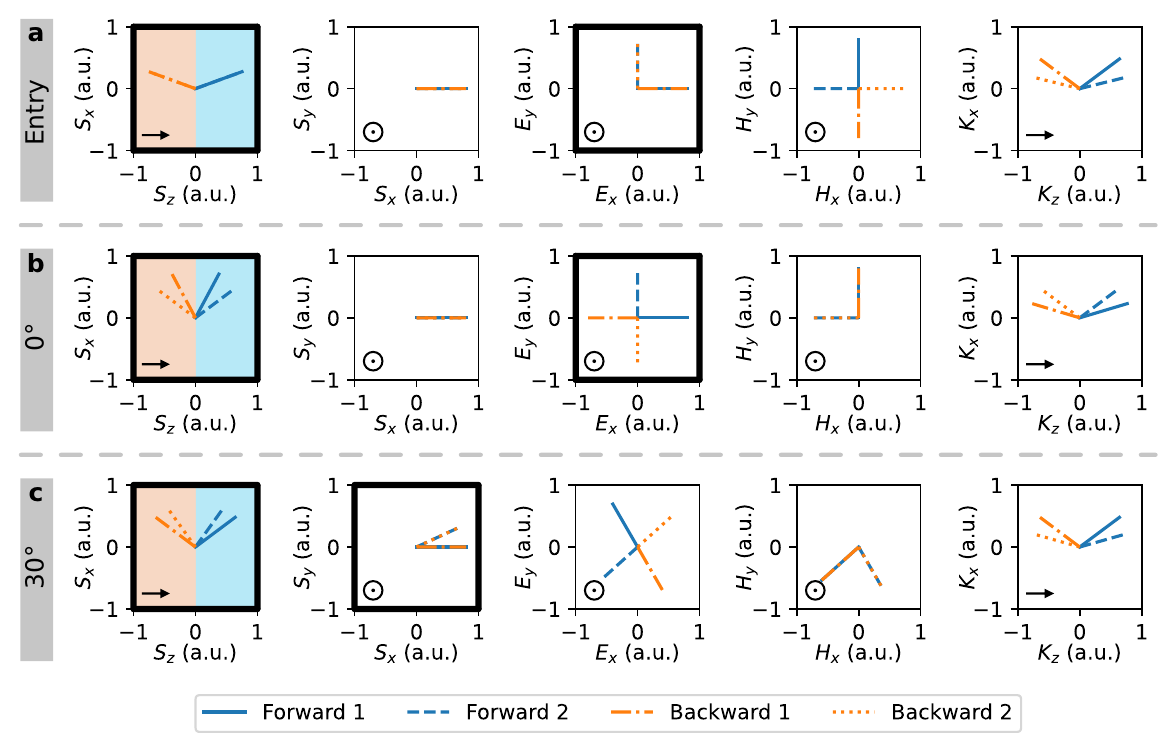}
	\caption{Schematic of the analysis of the partial waves (eigenvalues $K_{j, z}$ and eigenvectors $[E_{j, x}, H_{j, y}, E_{j, y}, -H_{j, x}]$). a. An isotropic entry half-space. b. A uniaxial crystal (anisotropic) whose principal axes are oriented along the laboratory axes $x$ and $y$. c. A uniaxial crystal whose principal axes are oriented at an arbitrary angle from the laboratory axes; the rotation angle of the crystal is $\ang{30}$ here.}
	\label{fig:eigen_sorting}
\end{figure}

Figure~\ref{fig:eigen_sorting} illustrates the analysis and sorting of the partial waves in three particular cases (all encountered when dealing with a cholesteric structure): an isotropic layer, a uniaxial crystal (anisotropic) whose principal axes are oriented along the laboratory axes $x$ and $y$, and a uniaxial crystal whose axes are not aligned with the $x$ and $y$ axes. First, the analysis of the direction of the Poynting vector $S_{j,z}$ (left column) enables to sort between forward (plotted in blue) and backward (plotted in orange) partial waves. Note that this is equivalent to sorting according to the direction of the wavevector $K_{j,z}$ (fifth column). Second, the direction of the electric field (third column) or of the Poynting vector (second column) along the $x$ or $y$ axes enables to identify the partial waves individually (plotted with different dashes). In each plot, the positive direction of the $z$ axis is displayed in the bottom-left corner for clarity. Each segment has been normalised to the same length. The elements that are effectively used to sort the partial waves have a thick black border and the others are displayed to provide a more complete picture of the situation.

\begin{figure}
  \centering
    \includegraphics[width=\textwidth]{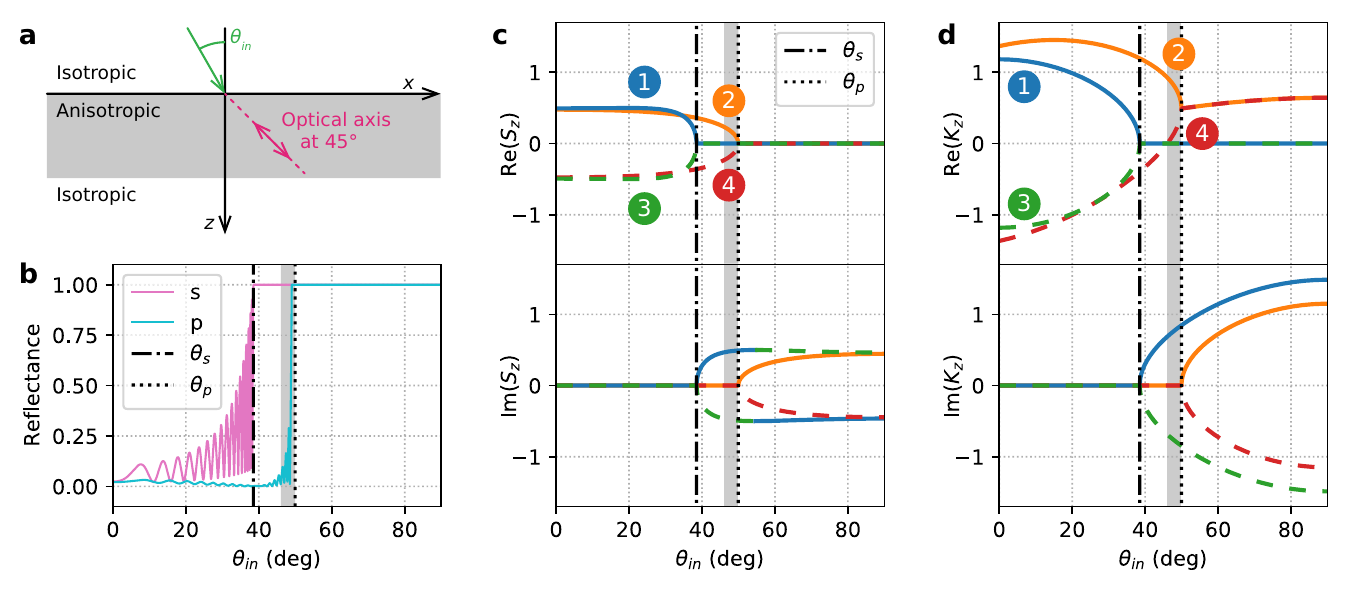}
	\caption{Illustration of the necessity to sort forward- and backward-propagating partial waves according to the direction of the Poynting vector.  a) Schematic of the situation. b) Reflectance plots for p and s polarisations and associated critical angles of total reflection according to Ref.~\cite{Riviere1978} (vertical dashed lines). c, d) Below the critical angle for p-polarised waves (gray shaded area), the Poynting vector (describing the direction of the energy flux) and the wavevector (describing the direction of the wave front) have opposite signs.}
	\label{fig:sort_kz}
\end{figure}

The wavevector $K_{j,z}$ and the Poynting vector $\mathcal{S}_{j,z}$ offer different information on the partial waves, indicating respectively the directions of the wave and of the energy flux. This is illustrated in Fig.~\ref{fig:sort_kz}, where we consider a slab of uniaxial crystal whose director lays in the $(xz)$ plane (see Figure~\ref{fig:sort_kz}a). In such situations, p- and s-polarised incident waves can have different angles of total reflection (see Figure~\ref{fig:sort_kz}b). Rivi\`{e}re provides analytical formulas for these critical angles~\cite{Riviere1978}, reported here as vertical dashed lines. In particular, when the optical axis of the crystal lays at $\ang{45}$ in the $(xz)$-plane, the critical angle of total reflection for p-polarisation is $\ang{49.9}$. Interestingly, at slightly smaller angles (gray shaded area on Fig.~\ref{fig:sort_kz}c-d), the partial wave $j=4$ exhibits real wavevector and Poynting vector of opposite signs, $\text{Re} [K_z]>0$ and $\text{Re}[\mathcal{S}_z]<0$. This has no effect on the predictions of the matrix methods because the imaginary part of the corresponding eigenvalue is zero.

\section{Discrete model for cholesterics}
\label{sec:cholesteric}

\begin{figure}
  \centering
    \includegraphics[width=0.5\textwidth]{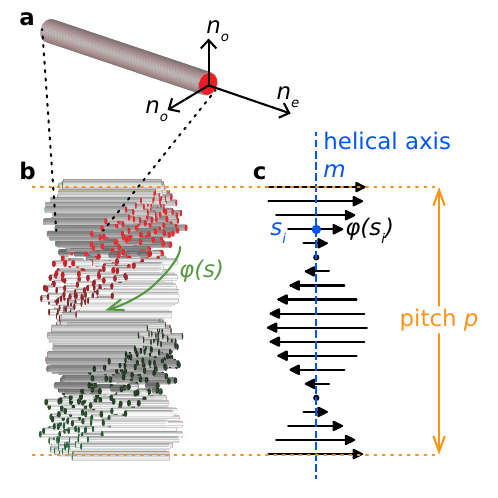}
	\caption{Schematic representation of a cholesteric structure in the $(xz)$ plane. The cholesteric structure repeats itself every half-pitch because the directors oriented towards $\ang{0}$ and $\ang{180}$ are equivalent; the red and green side of the elongated units and the tips of the arrows are a guide for the eye. a) Refractive indices of a birefringent unit. b) The material’s periodic twisted structure. c) A physical model consists in planes whose directors rotate around the helical axis $m$.}
	\label{fig:cholesteric}
\end{figure}

A cholesteric structure is an assembly of elongated birefringent units organised in a stack whose planes twist continuously around a helical axis at a periodicity called the pitch. At a given depth $s$ along the helical axis $m$, the elongated units point on average towards a preferred direction $n(s)$, which corresponds to a rotation angle $\phi(s)$ around the helical axis~\cite{DeVries1951, Frka-Petesic2019}, even if the direction of individual units may differ. Schematics of the birefringent units is displayed on Figure~\ref{fig:cholesteric}a (single unit) and b (assembly of units) while Figure~\ref{fig:cholesteric}c shows the directions of the rotating planes.

Optically, a cholesteric structure corresponds to a permittivity tensor $\epsilon(\phi(s))$ rotating at an angle $\phi(s)$ around the helical axis~\cite{Oseen1933, Belyakov1979}, where $\epsilon(0)$ corresponds to a birefringent unit aligned with the plane of incidence $(xz)$: $\epsilon_0 = \text{diag}(n_e^2, \: n_o^2, \: n_o^2)$ where $n_e$ and $n_o$ are the refractive indices of a birefringent unit. The optical response of the structure therefore directly relates to the periodic modulation of directors around the helical axis.

Under normal incidence of light, cholesterics reflect perfectly circularly polarised light of the handedness that matches the rotation of the helicoid (contrary to a mirror that swaps the polarisation) and in a narrow wavelength region. The central wavelength $\lambda_\text{B}$ of the reflection peak obeys Bragg’s law and is proportional to the cholesteric’s pitch $p$ and its average refractive index $n_\text{av}$ such that $\lambda_\text{B} = n_\text{av} p$~\cite{DeVries1951}. Under oblique incidence of light $\theta_{\text{in}}$, the reflection peak shifts such that $\lambda_\text{Bragg} = n_\text{av} p \cos(\theta_{\text{in}})$ and the structure is less polarisation-selective than at normal incidence~\cite{DeVries1951, Belyakov1979}.

Cholesteric structures that are distorted also reflect both polarisations~\cite{Kragt2019, Frka-Petesic2019}. Recently, Frka-Petesic \textit{et al.} developed a physical model to physically represent cholesteric structures, as well as the perturbation it encounters upon external constraints that affect the periodic directors and modify the optical response~\cite{Frka-Petesic2019}.

The optics of cholesterics is not trivial and no analytical formula can be found in the literature to predict the reflectance of a distorted cholesteric, for instance.

For modelling purposes, these continuous non-homogeneous materials are assimilated to discrete stacks of slices where each slice $i$ (at depth $s_i$ on the helical axis) possesses a director $n_i$ and rotation angle $\phi_i$, leading to a permittivity tensor $\epsilon(\phi_i)$, see Fig.~\ref{fig:cholesteric}c.

\section{Parameters for the figures}


\subsection{Figure~\ref{fig:validation_isotropic} (validation against analytical methods)}

For Fig.~\ref{fig:validation_isotropic}b, the parameters used to construct the Bragg stack (with \texttt{StackModel}) are

\begin{center}
\begin{tabular}{ |l|l|l| } 
\hline
\textbf{Parameter} & \textbf{Variable} & \textbf{Value} \\ \hline
Permittivity of the 1st layer & \texttt{eps\_list[0]} & $\begin{bmatrix*} 2.2^2 & 0 & 0 \\ 0 & 2.2^2 & 0 \\ 0 & 0 & 2.2^2\end{bmatrix*}$ \\ \hline
Permittivity of the 2nd layer & \texttt{eps\_list[1]} & $\begin{bmatrix*} 1.0 & 0 & 0 \\ 0 & 1.0 & 0 \\ 0 & 0 & 1.0\end{bmatrix*}$ \\ \hline
Thickness of the 1st layer (nm) & \texttt{thickness\_nm\_list[0]} & 200 \\ \hline
Thickness of the 2nd layer (nm) & \texttt{thickness\_nm\_list[1]} & 500 \\ \hline
Number of periods & \texttt{N\_per} & 10 \\ \hline
Index of entry medium & \texttt{n\_entry} & 1.0 \\ \hline
Index of exit medium & \texttt{n\_exit} & 2.2 \\ \hline
Angle of incidence (rad) & \texttt{theta\_in\_rad} & $\pi/3$ \\ \hline
Wavelength (nm) & \texttt{wl\_nm} & 400 to 800 \\ \hline
\end{tabular}
\end{center}

For Fig.~\ref{fig:validation_isotropic}d, the refractive indices and thicknesses used to construct the Bragg stack (with \texttt{StackModel}) are calculated from Eq.~\eqref{eq:yeh_exponential} with the number of periods indicated on Fig.~\ref{fig:validation_isotropic}c and $n_0=1$ and $n_s=4$.

The partial waves shown on Fig.~\ref{fig:validation_isotropic}e and f are extracted from a cholesteric structure. The parameters used to construct the cholesteric object (with \texttt{Cholesteric}) are
\begin{center}
\begin{tabular}{ |l|l|l| } 
\hline
\textbf{Parameter} & \textbf{Variable} & \textbf{Value} \\ \hline
Tilt (rad) & \texttt{tilt\_rad} & 0 \\ \hline
Pitch (nm) & \texttt{pitch360} & 500 \\\hline
Handedness & \texttt{handedness} & \texttt{1} (right) \\\hline
Slices per pitch & \texttt{resolution} & 360 \\\hline
\end{tabular}
\end{center}
The parameters used to construct the cholesteric model (with \texttt{CholestericModel}) are
\begin{center}
\begin{tabular}{ |l|l|l| } 
\hline
\textbf{Parameter} & \textbf{Variable} & \textbf{Value} \\ \hline
Number of periods & \texttt{N\_per} & 1 \\ \hline
Extraordinary index (along $x$) & \texttt{n\_e} & 1.4505 \\ \hline
Ordinary index (along $y$ and $z$) & \texttt{n\_o} & 1.4155 \\ \hline
Index of entry medium & \texttt{n\_entry} & 1.433 \\ \hline
Index of exit medium & \texttt{n\_exit} & 1.433 \\ \hline
Angle of incidence (rad) & \texttt{theta\_in\_rad} & 0 (e) and $\pi/3$ (f) \\ \hline
Wavelength (nm) & \texttt{wl\_nm} & $\lambda_{Bragg} =$ e) $716.5$ f) $358.25$ \\ \hline
\end{tabular}
\end{center}
The partial waves are extracted from the 15th anisotropic slice and b) the 30th anisotropic slice that have a rotation angle of $\ang{15}$ and $\ang{30}$ around the $z$ axis, respectively.

\subsection{Figure~\ref{fig:showcase} (various examples of cholesterics)}

For Fig.~\ref{fig:showcase}a (right-handed), the parameters used to construct the cholesteric object (with \texttt{Cholesteric}) are
\begin{center}
\begin{tabular}{ |l|l|l| } 
\hline
\textbf{Parameter} & \textbf{Variable} & \textbf{Value} \\ \hline
Tilt (rad) & \texttt{tilt\_rad} & 0 \\ \hline
Pitch (nm) & \texttt{pitch360} & 500 \\\hline
Handedness & \texttt{handedness} & right (\texttt{1}) \\\hline
Slices per pitch & \texttt{resolution} & 40 \\\hline
Compression ratio & \texttt{alpha} & 1 \\\hline
Arbitrary distortion & \texttt{d} & none (\texttt{1}) \\\hline
\end{tabular}
\end{center}
The parameters used to construct the cholesteric model (with \texttt{CholestericModel}) are
\begin{center}
\begin{tabular}{ |l|l|l| } 
\hline
\textbf{Parameter} & \textbf{Variable} & \textbf{Value} \\ \hline
Number of periods & \texttt{N\_per} & 10 \\ \hline
Extraordinary index (along $x$) & \texttt{n\_e} & 1.533 \\ \hline
Ordinary index (along $y$ and $z$) & \texttt{n\_o} & 1.333 \\ \hline
Index of entry medium & \texttt{n\_entry} & 1.433 \\ \hline
Index of exit medium & \texttt{n\_exit} & 1.433 \\ \hline
Angle of incidence (rad) & \texttt{theta\_in\_rad} & 0 \\ \hline
Wavelength (nm) & \texttt{wl\_nm} & 400 to 800 \\ \hline
\end{tabular}
\end{center}

For Fig.~\ref{fig:showcase}b (left-handed), the parameters used to construct the cholesteric object (with \texttt{Cholesteric}) are the same as Fig.~\ref{fig:showcase}a except
\begin{center}
\begin{tabular}{ |l|l|l| } 
\hline
\textbf{Parameter} & \textbf{Variable} & \textbf{Value} \\ \hline
Handedness & \texttt{handedness} & left (\texttt{-1}) \\\hline
\end{tabular}
\end{center}
The parameters used to construct the cholesteric model (with \texttt{CholestericModel}) are the same as Figure~\ref{fig:showcase}a.

For Fig.~\ref{fig:showcase}c (tilted and normal incidence), the parameters used to construct the cholesteric object (with \texttt{Cholesteric}) are the same as Fig.~\ref{fig:showcase}a except:
\begin{center}
\begin{tabular}{ |l|l|l| } 
\hline
\textbf{Parameter} & \textbf{Variable} & \textbf{Value} \\ \hline
Tilt (rad) & \texttt{tilt\_rad} & $\pi/6$ \\ \hline
\end{tabular}
\end{center}
The parameters used to construct the cholesteric model (with \texttt{CholestericModel}) are the same as Figure~\ref{fig:showcase}a.

For Fig.~\ref{fig:showcase}d (tilted and oblique incidence), the parameters used to construct the cholesteric object (with \texttt{Cholesteric}) are the same as Fig.~\ref{fig:showcase}a except
\begin{center}
\begin{tabular}{ |l|l|l| } 
\hline
\textbf{Parameter} & \textbf{Variable} & \textbf{Value} \\ \hline
Tilt (rad) & \texttt{tilt\_rad} & $\pi/6$ \\ \hline
\end{tabular}
\end{center}
The parameters used to construct the cholesteric model (with \texttt{CholestericModel}) are the same as Fig.~\ref{fig:showcase}a except:
\begin{center}
\begin{tabular}{ |l|l|l| } 
\hline
\textbf{Parameter} & \textbf{Variable} & \textbf{Value} \\ \hline
Angle of incidence (rad) & \texttt{theta\_in\_rad} & $70 \pi/180$ \\ \hline
\end{tabular}
\end{center}

For Fig.~\ref{fig:showcase}e (compressed), the parameters used to construct the cholesteric object (with \texttt{Cholesteric}) are the same as Fig.~\ref{fig:showcase}a except:
\begin{center}
\begin{tabular}{ |l|l|l| } 
\hline
\textbf{Parameter} & \textbf{Variable} & \textbf{Value} \\ \hline
Compression ratio & \texttt{alpha} & 0.7 \\\hline
\end{tabular}
\end{center}
The parameters used to construct the cholesteric model (with \texttt{CholestericModel}) are the same as Fig.~\ref{fig:showcase}a.

For Fig.~\ref{fig:showcase}f (distorted), the parameters used to construct the cholesteric object (with \texttt{Cholesteric}) are the same as Fig.~\ref{fig:showcase}a except:
\begin{center}
\begin{tabular}{ |l|l|l| } 
\hline
\textbf{Parameter} & \textbf{Variable} & \textbf{Value} \\ \hline
Arbitrary distortion & \texttt{d} & 3 \\\hline
\end{tabular}
\end{center}
The parameters used to construct the cholesteric model (with \texttt{CholestericModel}) are the same as Figure~\ref{fig:showcase}a.

\subsection{Figure~\ref{fig:mixed_model} (\texttt{MixedModel} example)}

The parameters used to construct the cholesteric objects \texttt{chole\_1} and \texttt{chole\_2} (with \texttt{Cholesteric}) are
\begin{center}
\begin{tabular}{ |l|l|l| } 
\hline
\textbf{Parameter} & \textbf{Variable} & \textbf{Value} \\ \hline
Tilt (rad) & \texttt{tilt\_rad} & 0 \\ \hline
Pitch (nm) & \texttt{pitch360} & 550 \\\hline
Handedness & \texttt{handedness} & right (\texttt{1}) and left (\texttt{-1}) \\\hline
Slices per pitch & \texttt{resolution} & 40 \\\hline
\end{tabular}
\end{center}
The parameters used to construct the corresponding cholesteric models (with \texttt{CholestericModel}) are
\begin{center}
\begin{tabular}{ |l|l|l| } 
\hline
\textbf{Parameter} & \textbf{Variable} & \textbf{Value} \\ \hline
Number of periods & \texttt{N\_per} & 30 \\ \hline
Extraordinary index (along $x$) & \texttt{n\_e} & 1.443 \\ \hline
Ordinary index (along $y$ and $z$) & \texttt{n\_o} & 1.423 \\ \hline
Index of entry medium & \texttt{n\_entry} & 1.433 \\ \hline
Index of exit medium & \texttt{n\_exit} & 1.433 \\ \hline
Angle of incidence (rad) & \texttt{theta\_in\_rad} & $40 \pi/180$ \\ \hline
Wavelength (nm) & \texttt{wl\_nm} & 400 to 800 \\ \hline
\end{tabular}
\end{center}

\subsection{Figure~\ref{fig:performances} (performances of the SM and TM methods)}

The matrix elements shown on Fig.~\ref{fig:performances}a are calculated from a cholesteric structure. The spectra shown on Fig.~\ref{fig:performances}c and d are calculated from a cholesteric structure with similar parameters. The colormap shown on Fig.~\ref{fig:performances}b is made by further varying the birefringence and the number of periods of this cholesteric.

The parameters used to construct the cholesteric object (with \texttt{Cholesteric}) are:

\begin{center}
\begin{tabular}{ |l|l|l| } 
\hline
\textbf{Parameter} & \textbf{Variable} & \textbf{Value} \\ \hline
Tilt (rad) & \texttt{tilt\_rad} & 0 \\ \hline
Pitch (nm) & \texttt{pitch360} & 500 \\\hline
Handedness & \texttt{handedness} & right (\texttt{1}) \\\hline
Slices per pitch & \texttt{resolution} & a) 20, b), c), d) 40 \\\hline
\end{tabular}
\end{center}

The parameters used to construct the cholesteric model (with \texttt{pyllama.CholestericModel}) are:

\begin{center}
\begin{tabular}{ |l|l|l| } 
\hline
\textbf{Parameter} & \textbf{Variable} & \textbf{Value} \\ \hline
Number of periods & \texttt{N\_per} & See figure \\ \hline
Extraordinary index (along $x$) & \texttt{n\_e} & See figure \\ \hline
Ordinary index (along $y$ and $z$) & \texttt{n\_o} & See figure \\ \hline
Average index &  & 1.433 \\ \hline
Birefringence &  & Variable \\ \hline
Index of entry medium & \texttt{n\_entry} & 1.433 \\ \hline
Index of exit medium & \texttt{n\_exit} & 1.433 \\ \hline
Angle of incidence (rad) & \texttt{theta\_in\_rad} & a) 0, b, c, d) $\pi/4$ \\ \hline
Wavelength (nm) & \texttt{wl\_nm} & a) $\lambda_{Bragg} = 716.5$ \\ \hline
\end{tabular}
\end{center}

\subsection{Figure~\ref{fig:eigen_sorting} (sorting of the partial waves, examples)}

The partial waves are extracted from a cholesteric structure. The parameters used to construct the cholesteric object (with \texttt{Cholesteric}) are:

\begin{center}
\begin{tabular}{ |l|l|l| } 
\hline
\textbf{Parameter} & \textbf{Variable} & \textbf{Value} \\ \hline
Tilt (rad) & \texttt{tilt\_rad} & 0 \\ \hline
Pitch (nm) & \texttt{pitch360} & 1000 \\\hline
Handedness & \texttt{handedness} & right (\texttt{1}) \\\hline
Slices per pitch & \texttt{resolution} & 360 \\\hline
\end{tabular}
\end{center}

The parameters used to construct the cholesteric model (with \texttt{CholestericModel}) are:

\begin{center}
\begin{tabular}{ |l|l|l| } 
\hline
\textbf{Parameter} & \textbf{Variable} & \textbf{Value} \\ \hline
Number of periods & \texttt{N\_per} & 1 \\ \hline
Extraordinary index (along $x$) & \texttt{n\_e} & 3 \\ \hline
Ordinary index (along $y$ and $z$) & \texttt{n\_o} & 1.2 \\ \hline
Index of entry medium & \texttt{n\_entry} & 2.1 \\ \hline
Index of exit medium & \texttt{n\_exit} & 2.1 \\ \hline
Angle of incidence (rad) & \texttt{theta\_in\_rad} & $\pi/9$ \\ \hline
Wavelength (nm) & \texttt{wl\_nm} & $\lambda_{Bragg} \approx 1346$ \\ \hline
\end{tabular}
\end{center}

The partial waves are extracted from a) the entry isotropic half-space, b) the first anisotropic slice and c) the 30th anisotropic slice that has a rotation angle of $\ang{30}$ around the $z$ axis.

\subsection{Figure~\ref{fig:sort_kz} (sorting of the partial waves with the wavevector or the Poynting vector)}

The parameters used to construct the slab (with \texttt{SlabModel}) are:
\begin{center}
\begin{tabular}{ |l|l|l| } 
\hline
\textbf{Parameter} & \textbf{Variable} & \textbf{Value} \\ \hline
Permittivity & \texttt{eps} & $\begin{bmatrix*} 1.682^2 & 0 & 0 \\ 0 & 1.183^2 & 0 \\ 0 & 0 & 1.183^2\end{bmatrix*}$ \\ \hline
Thickness (nm) & \texttt{thickness\_nm} & 4000 \\ \hline
Rotation angle (rad) & \texttt{rotangle\_rad} & $\pi/8$ \\ \hline
Rotation axis & \texttt{rotaxis} & $y$ \\ \hline
Index of entry medium & \texttt{n\_entry} & 1.9 \\ \hline
Index of exit medium & \texttt{n\_exit} & 1.433 \\ \hline
Angle of incidence (rad) & \texttt{theta\_in\_rad} & 0 to $\pi/4$ \\ \hline
Wavelength (nm) & \texttt{wl\_nm} & 500 \\ \hline
\end{tabular}
\end{center}
Figure \ref{fig:sort_kz}b represents the partial waves extracted for an incident angle of $\ang{49.9}$.






\end{document}